\documentclass[aps,amsmath,amssymb,preprintnumbers,nofootinbib,11pt]{revtex4}
\usepackage[utf8]{inputenc} %utf8
\usepackage{graphicx}
\graphicspath{{./figures/}}
\usepackage{url}
\usepackage[bookmarks, pagebackref=false]{hyperref}
\usepackage[usenames,dvipsnames]{xcolor}

\usepackage{makecell}

\definecolor{orange}{cmyk}{0,0.5,1,0}
\definecolor{rossoCP3}{cmyk}{0,.88,.77,.40}
\definecolor{graa}{rgb}{0.8,0.8,0.8}
\definecolor{blaa}{rgb}{0.2,0.2,0.6}
	\hypersetup{
		bookmarksopen,
			bookmarksnumbered,
			citecolor=blaa, 		%color of links to bibliography
			linkcolor=rossoCP3,	%color of internal links
			urlcolor=rossoCP3,			%color of external links
			}
\usepackage{amsthm}
\usepackage{bm}% bold math
\usepackage{bbm}
\usepackage{pxfonts}
\usepackage{amsmath,amssymb,amsfonts}
\usepackage{color}
\usepackage{float}
\usepackage{hyperref}
\usepackage[Symbolsmallscale]{upgreek}
\usepackage{amsmath}
\usepackage{amsfonts}
\usepackage{amssymb,dsfont}
\usepackage{graphicx}
\usepackage{amssymb}
\usepackage[vcentermath]{youngtab}
\usepackage[all]{xy}
\usepackage{pstricks}
\usepackage{dsfont}%
\usepackage{bbold}
\usepackage[normalem]{ulem}

\usepackage{placeins}
\usepackage{xspace}
\usepackage{cancel} % to make the barred text notation

\usepackage{slashed}
\usepackage{natbib}

\usepackage{feynmf}

\usepackage{braket}
\usepackage{slashed}

\newcommand{\ee}{\end{equation}}
\newcommand{\be}{\begin{equation}}
\newcommand{\bea}{\begin{align}}
\newcommand{\eea}{\end{alig}}
\newcommand{\nn}{~\nonumber \\}

\newcommand{\bmp}{\noindent\begin{minipage}{16cm}}
\newcommand{\emp}{\end{minipage}\vskip 7mm} % 7mm untightened

\newcommand   \cO {\mathcal{O}}

\newcommand{\Tr}{\text{Tr}}

\newcommand{\eps}  {\epsilon}

\def\lsim{\mathrel{\rlap{\lower4pt\hbox{\hskip1pt$\sim$}}
    \raise1pt\hbox{$<$}}}                % less than or approx. symbol
\def\gsim{\mathrel{\rlap{\lower4pt\hbox{\hskip1pt$\sim$}}
    \raise1pt\hbox{$>$}}}                % greater than or approx. symbol

\begin{document}
%%%%%%%%%%%%%%%%%%%%%%%%%%%%%%%%%%%%%%%%%%%%%%%%%%%%%%%%%%%%%%%%%%%%%%%%%%%

\title{The analytic structure of the fixed charge expansion}
\author{Oleg {\sc Antipin}$^{\color{rossoCP3}{\clubsuit}}$}
\email{oantipin@irb.hr}
\author{Jahmall {\sc Bersini}
$^{\color{rossoCP3}{\clubsuit}}$}
\email{jbersini@irb.hr}
\author{Francesco {\sc Sannino} $^{\color{rossoCP3} {\diamondsuit},\color{rossoCP3}{{\varheartsuit},\heartsuit,\vardiamondsuit}}$}
\email{sannino@cp3.dias.sdu.dk}
\author{Matías Torres $^{\color{rossoCP3}{\heartsuit}}$}
\email{matiasignacio.torressandoval@unina.it}

\affiliation{{ $^{\color{rossoCP3}{\clubsuit}}$ Rudjer Boskovic Institute, Division of Theoretical Physics, Bijeni\v cka 54, 10000 Zagreb, Croatia}\\{ $^{\color{rossoCP3}{\diamondsuit}}$\color{rossoCP3} {CP}$^{ \bf 3}${-Origins}} \& the Danish Institute for Advanced Study {\color{rossoCP3}\rm{Danish IAS}},  University of Southern Denmark, Campusvej 55, DK-5230 Odense M, Denmark. \\
\mbox{ $^{\color{rossoCP3}{\varheartsuit}}$ Scuola Superiore Meridionale, \\ Largo S. Marcellino, 10, 80138 Napoli, Italy}
\mbox{  $^{\color{rossoCP3}{\heartsuit}}$ Dipartimento di Fisica, E. Pancini, Universit\'a di Napoli, Federico II, INFN sezione di Napoli.} \\ \mbox{Complesso Universitario di Monte S. Angelo Edificio 6, via Cintia, 80126 Napoli, Italy.} \\
\mbox{$^{\color{rossoCP3}{\vardiamondsuit}}$  CERN, Theoretical Physics Department, 1211 Geneva 23, Switzerland.}}

\begin{abstract}
{
We investigate the analytic properties of the fixed charge expansion for a number of conformal field theories in different space-time dimensions. The models investigated here are $O(N)$ and $QED_3$. We show that in $d=3-\epsilon$ dimensions the contribution to the $O(N)$ fixed charge $Q$ conformal dimensions obtained in the double scaling limit of large charge and vanishing $\epsilon$ is non-Borel summable, doubly factorial divergent, and with order $\sqrt{Q}$ optimal truncation order. By using resurgence techniques we show that the singularities in the Borel plane are related to worldline instantons that were discovered in the other double scaling limit of large $Q$ and $N$ of Ref.~\cite{Dondi:2021buw}. In $d=4-\epsilon$ dimensions the story changes since in the same large $Q$ and small $\epsilon$ regime the next order corrections to the scaling dimensions lead to a convergent series. The resummed series displays a new branch cut singularity which is relevant for the stability of the $O(N)$ large charge sector for negative $\epsilon$. Although the $QED_3$ model shares the same large charge behaviour of the $O(N)$ model, we discover that at leading order in the large number of matter field expansion the large charge scaling dimensions are Borel summable, single factorial divergent, and with order $Q$ optimal truncation order.}
  \\~\\
{\footnotesize  \it Preprint: RBI-ThPhys-2022-6}
\end{abstract}

\maketitle

\small
\newpage

\section{Introduction} 

Our understanding of Nature is seriously hampered by our limited knowledge of quantum field theory  (QFT) in the strongly coupled regime. Time-honoured examples range from Quantum Chromodynamics (QCD) to the understanding of critical dynamics relevant for a plethora of physical applications from condensed matter physics \cite{Cardy:1996xt} to epidemiology \cite{DellaMorte:2020wlc,10.3389/fams.2021.659580,cacciapaglia2021epidemiological,cacciapaglia2020second,Cacciapaglia:2021vvu}. 
Several tools have been developed to tackle strongly coupled dynamics including using weakly coupled expansions to deduce non-perturbative information, see for example \cite{LeGuillou:1990nq} for a review. Being, in general, the perturbative series asymptotic, the non-perturbative information is expected to be contained in the analytic structure of their Borel transform. A well-known example is given by instantons singularities in the Borel plane, which have a semiclassical interpretation in terms of non-trivial classical trajectories \cite{Lipatov:1976ny}. The mathematical framework that systematizes the idea of inferring non-perturbative physics from perturbation theory was developed by J. Ecalle in the $'80$s \cite{Ecalle} and takes the name of \emph{resurgence theory} \footnote{See also \cite{Dorigoni:2014hea, Aniceto:2018bis} for a physics-oriented review.}. In the last years, many works have successfully applied these ideas to QFT \cite{Dunne:2012ae, Dunne:2012zk, Argyres:2012ka, Cherman:2014ofa, Dorigoni:2015dha, Behtash:2015kna, Yamazaki:2017ulc, Boito:2017cnp, Marino:2019eym, Ishikawa:2019tnw, Ishikawa:2019oga, Bersini:2019axn, Abbott:2020qnl, Borinsky:2020vae, Dunne:2021acr, Maiezza:2021mry, Marino:2021dzn, Argyres:2012vv, Dunne:2021lie}
providing a novel perspective on various non-perturbative phenomena such as renormalons, instantons, and quark-hadron duality. Moreover,   an intimate relation has emerged between resurgence and phase transitions when the latter are seen as Stokes phenomena \cite{Pisani, Kanazawa:2014qma, Buividovich:2015oju, Ahmed:2017lhl, Fujimori:2021oqg, Basar:2021gyi}.

At the same time, an independent line of research is being developed and it is aimed at the understanding of the strongly coupled regime of conformal field theories (CFT)s via a large-charge induced semiclassical expansion  \cite{Hellerman:2015nra}. Here one uses EFT methods \cite{Monin:2016jmo, Alvarez-Gaume:2016vff, Banerjee:2017fcx, Hellerman:2017sur,  delaFuente:2018qwv, Orlando:2019hte, Banerjee:2019jpw, Cuomo:2020rgt, Cuomo:2021ygt, Hellerman:2021yqz, Cuomo:2021cnb, Pellizzani:2021hzx, Hellerman:2021qzz, Banerjee:2021bbw}. In fact, semiclassical expansions have shown to be useful even in resumming infinite series of Feynman diagrams thereby helping shed light on higher order computations in different regimes \cite{Alvarez-Gaume:2019biu, Badel:2019oxl, Badel:2019khk, Antipin:2020abu, Antipin:2020rdw, Antipin:2021akb, Antipin:2021jiw, Arias-Tamargo:2019xld, Arias-Tamargo:2020fow, Jack:2020wvs, Jack:2021ypd, Jack:2021lja, Jack:2021ziq, Giombi:2020enj, Giombi:2021zfb, Araujo:2021sjv, Rodriguez-Gomez:2022gbz}. The approach can be extended to non-conformal QFTs as illustrated in \cite{Son:1995wz, Orlando:2019skh, Orlando:2020yii, Moser:2021bes, Orlando:2021usz}, with possible physical applications such as the study of multi-boson production processes in the Standard Model.

The potential effectiveness of the large-charge expansion for small values of the charge, that seems to emerge by comparing predictions to lattice results \cite{Banerjee:2017fcx, Banerjee:2019jpw}, partially motivated the first analysis implementing resurgence for the large-charge expansion  of \cite{Dondi:2021buw}. {An interesting investigation of the exponentially small corrections to the large R-charge expansion in $\mathcal{N}=2$ superconformal QCD appeared a few weeks later in \cite{Hellerman:2021duh}.  In \cite{Dondi:2021buw}, the authors considered the spectrum of charge $Q$ operators for the critical $O(N)$ model in $d=3$ dimensions in the double-scaling limit
\be
Q\to\infty \,, \qquad N\to\infty \,, \qquad \frac{Q}{N}  \text{~~~fixed} \,.
\ee
In this limit, the scaling dimensions of the lowest-lying operators with total charge $Q$ assume the form \cite{Alvarez-Gaume:2019biu} 
\be
\Delta_Q = \sum_{j=-1} \frac{1}{N^j}\Delta_j \left(\frac{Q}{N} \right) \,.
\label{doubleQN}
\ee
By expanding the $\Delta_j$ in the small $\frac{Q}{N}$ limit, one recovers the ordinary $1/N$ expansion \cite{Moshe:2003xn}, while for ${Q}/{N} \gg 1$, Eq.\eqref{doubleQN} reproduces the general form of the large-charge expansion in generic non-supersymmetric relativistic CFTs \footnote{When $d$ is even, one needs to include in the expansion $Q^p \log(Q)$ terms, with $p$ to be determined,  induced by the cancellation of UV divergences \cite{Cuomo:2020rgt}.} \cite{Hellerman:2015nra, Monin:2016jmo, Cuomo:2020rgt}
\begin{align}
\Delta_{Q}= Q^{\frac{d}{d-1}}\left[\alpha_{1}+ \alpha_{2} Q^{\frac{-2}{d-1}}+\alpha_3 Q^{\frac{-4}{d-1}}+\ldots\right] +Q^0\left[\beta_0+ \beta_{1} Q^{\frac{-2}{d-1}}+\ldots\right] + \cO\left( Q^{-\frac{d}{d-1}}\right)  \ ,
\label{largecharge}
\end{align}
which can be derived from the large-charge effective action without assuming the presence of other expansion parameters apart from $1/Q$. The central object studied in \cite{Dondi:2021buw} is the functional determinant of a free scalar field with mass equal to the chemical potential $\mu$  (conjugated to the fixed charge $Q$) on $\mathbb{R}\times S^{d-1}$ 
\be \label{labello}
\log \left( \det \left(-\partial_0^2 - \Delta_{S^{d-1}} + \mu^2   \right) \right) = \sum_{\ell=0} n_\ell \sqrt{J_\ell^2 + \mu^2} \,,
\ee
where $\Delta_{S^{d-1}}$ is the Laplacian on the $(d-1)$-sphere, whose eigenvalues $J^2_\ell$ and their multiplicity $n_\ell$ are given by
\be \label{eigen}
 J^2_\ell = \ell (\ell+d-2) \,, \qquad n_\ell=\frac{(2\ell+d-2) \Gamma(\ell+d-2)}{\Gamma(\ell+1) \Gamma(d-1)} \,.
\ee
In fact, for technical reasons, the theory is Weyl-mapped to $\mathbb{R}\times S^{d-1}$, and, via the state operator correspondence \cite{Cardy:1984rp,Cardy:1985lth}, the energy levels on the cylinder are linked to the associated spectrum of scaling dimensions. $\Delta_{-1}$ in Eq.\eqref{doubleQN} is then obtained as the Legendre transform of Eq.\eqref{labello} with respect to the chemical potential to express it in terms of the fixed charge $Q$.

The small $\frac{Q}{N}$ expansion of $\Delta_{-1}$ is convergent with a radius of convergence related to the appearance of a zero-mode in the spectrum. On the other hand, the large $\frac{Q}{N}$ expansion of $\Delta_{-1}$ diverges $(2n)!$ factorially, and its Borel transform exhibits an infinite number of singularities on the positive real axis which, according to resurgence theory \footnote{Notice that a priori is not known whether QFT observables satisfy the axiom of resurgence theory, i.e they are \emph{resurgent functions}. In this work, we \emph{assume} this condition. For a recent discussion on this point, including counterexamples, we refer the interested reader to \cite{DiPietro:2021yxb}.}, indicate the emergence of non-perturbative corrections. The leading non-perturbative contributions scale as $e^{-\sqrt{Q}}$ and stem from worldline instantons describing the geodesic motion of a free particle with mass $\mu$ moving on close trajectories \cite{Schubert:2001he}. 
% Since these corrections originate from the geometrical properties of the compactification manifold, the authors of \cite{Dondi:2021buw} conjectured the above to be a general property of the large-charge expansion. It was envisioned to be consequence of the effective action describing the large-charge sector of the three-dimensional $O(N)$ CFT that is by itself an asymptotic series. The resulting optimal truncation order, for any $N$, is $n_{\text{opt}} \approx \sqrt{Q}$ with the related error $\cO\left(e^{-\sqrt{Q}} \right)$. The $O(N)$ model in $d=5$ in the same double-scaling limit has been investigated in \cite{Moser:2021bes}, reaching similar conclusions.  
Since these corrections originate from the geometrical properties of the compactification manifold, the authors of \cite{Dondi:2021buw} conjectured the above to be a general property of the large-charge expansion in the three-dimensional $O(N)$  CFT. It was envisioned to be a consequence of the effective action describing the large-charge sector of the theory that is by itself an asymptotic series. The resulting optimal truncation order, for any $N$, is $n_{\text{opt}} \approx \sqrt{Q}$ with the related error $\cO\left(e^{-\sqrt{Q}} \right)$. The $O(N)$ model in $d=5$ in the same double-scaling limit has been investigated in \cite{Moser:2021bes}, reaching similar conclusions. 
Here we add information on the convergence properties of the large-charge expansion by addressing various models displaying very different large order behaviours. Along our journey, we will encounter convergent, asymptotic but Borel summable, and non-Borel summable series; in the first case we will investigate what one can learn on the physics of the
expansion from a finite number of coefficients. To this end, our main tool will be the Darboux's theorem \cite{Darboux, Darboux2}, which relates the behaviour of a function around its non-analytical points to the rate of growth of the coefficients of its series expansion around regular points.  Physical applications were explored in \cite{Darboux, Frazer:1961zz, Hunter:1973zz, Kazakov:1978ey, Fischer:1997bs,  Stephanov:2006dn, Caprini:2017ikn,Dondi:2019ivp, Costin:2020hwg,  Dondi:2020qfj}. 

We organize the work as follows. In Sec.\ref{3minuseps}, we consider the critical $g^2 (\phi_i\phi_i)^3$ theory in $d=3-\eps$, which has been investigated in \cite{Badel:2019khk, Jack:2020wvs} in the double-scaling limit 
\be \label{doubleQeps}
Q \to \infty \,, \qquad g \to 0 \,, \qquad g Q = \text{fixed} \,.
\ee
resulting in the following semiclassical expansion
\begin{equation} \label{TQ}
\Delta_{Q} = \sum_{j=-1}^{\infty} g^{*j}\Delta_j(g^*  Q) \ .
\end{equation}
We discover that the small-charge (i.e. the small $g Q$) expansions of $\Delta_{-1}$ and $\Delta_0$ are convergent and share the same radius of convergence. We observe that, as in \cite{Dondi:2021buw}, the leading singularity, which is an algebraic branch point, occurs when the mass of a certain mode vanishes. Moreover, the small $g Q$ expansion of $\Delta_0$ provides an interesting example of how the program of reconstructing the analytic structure of a function from a limited number of expansion coefficients can fail. In fact, even the precise identification of the radius of convergence requires more than one hundred expansion coefficients. However, we are able to make progress by identifying the source of the problem in the occurrence of two coincident singularities for which we can disentangle their contributions.

Additionally, $\Delta_0$ in Eq.\eqref{TQ} is the functional determinant of the fluctuations around the classical solution and, in $O(N)$-invariant theories in any $d$, receives contributions from three types of modes \cite{Alvarez-Gaume:2016vff, Antipin:2020abu}: one massless \emph{conformal mode}, one massive \emph{radial mode}, and $N-2$ \emph{spectator modes}. The computation of the large $g Q$ expansion of the full $\Delta_0$ is technically challenging and the approaches considered in the literature resorted to numerical fits \cite{delaFuente:2018qwv, Badel:2019oxl, Badel:2019khk, Jack:2020wvs, Jack:2021ypd} and numerical evaluation of integrals \cite{Cuomo:2021cnb}, in order to determine the first few coefficients. For the sake of simplicity, in this exploratory work we focus only on the contribution of the spectator modes, which is given by the functional determinant \eqref{labello} in $d=3$ i.e. it is exactly the same object considered in \cite{Dondi:2021buw}. At the same time, due to the different double-scaling limit considered, our large-charge expansion of Eq.\eqref{labello} differs from the one considered in \cite{Dondi:2021buw} but, not surprisingly, share all its features i.e. a $(2n)!$ factorial growth of the coefficients related to the same non-perturbative effects driven by worldline instantons. 
We are, therefore, able to confirm the results of \cite{Dondi:2021buw} in a different double-scaling limit, providing strong evidence for the non-perturbative corrections due to worldline instantons being a general feature of the large-charge expansion on $\mathbb{R}\times S^2$. 

Motivated by the geometrical origin of these non-perturbative corrections, in Sec.\ref{4minuseps}, we move to $\mathbb{R}\times S^{3-\eps}$ and study the $g (\phi_i \phi_i)^2$ $O(N)$ model in $d=4-\epsilon$ dimensions in the double-scaling limit \eqref{doubleQeps}, which has been previously considered in \cite{Badel:2019oxl, Antipin:2020abu, Jack:2021ypd}. In particular, an interesting diagrammatic argument that links the large order behaviour of the coefficients of the small $g Q$ expansion of $\Delta_j$ for different $j$ has been given in \cite{Badel:2019oxl}. After elaborating the consequences of this proposal for the analytical structure of the $\Delta_j$, we show that the small $g Q$ expansion of $\Delta_0$ contradicts it; the situation is completely analogous to the $d=3-\eps$ case; the small $g Q$ expansion of both $\Delta_{-1}$ and $\Delta_0$ is convergent, with the radial mode becoming massless at the leading singular point. Moreover, $\Delta_0$ features two coincident leading singularities. The branch point determining the radius of convergence lies on the negative $g Q$ axis and it is, therefore, possible to smoothly continue the small-charge expansion to large positive values of the charge. On the other hand, this branch point is related to the instability of the large-charge sector of the (metastable) ultraviolet FP of the quartic $O(N)$ theory in $4<d<6$, which has been recently pointed out in \cite{Giombi:2020enj, Antipin:2021jiw} and related to a phase transition on the cylinder in \cite{Moser:2021bes}. In fact, this FP can be reached by continuing the $\eps$-expansion to negative values of $\eps$ \cite{Giombi:2014iua} and occurs at negative $g$. Interestingly, the type of the leading singularities in both $\Delta_{-1}$ and $\Delta_0$ is exactly the same in the $d=4-\eps$ and $d=3-\eps$ cases. Motivated by this observation we make a slight detour and study the small-charge expansion in the cubic $O(N)$ model in $d=6-\eps$ and the $U(N)\times U(M)$ model in $d=4-\epsilon$ at the leading order of the semiclassical expansion \eqref{TQ}. Intriguingly, we discover that the structure of the leading singularity at the leading order of the semiclassical expansion is shared among all these theories, despite the differences in $d$, matter content, and symmetries. 

In order to study the large $g Q$ expansion of $\Delta_0$, we consider the contribution of the spectator modes, leaving the remaining two modes for future work. As opposed to the three-dimensional $O(N)$ theory, we discover that the large $g Q$  expansion of these contributions converges. In the case of the spectator modes, this is traced back to the convergence of the heat kernel expansion on odd-dimensional spheres \cite{Camporesi:1990wm}. Moreover, we are able to resum the large $g Q$ expansion of the contribution of the spectator modes obtaining a simple analytical expression not involving infinite sums. Its analytical structure reveals a previously unnoticed branch cut on the negative $g Q$ axis starting at $g Q = 0$, which makes the fixed-charge sector of the $O(N)$ theory in $4<d<6$ unstable for \emph{any} value of the charge. This differs from previous investigations \cite{Giombi:2020enj, Antipin:2021jiw}, which observed such instabilities only above a critical (finite) value of the charge.

In Sec.\ref{sez4}, we return to $\mathbb{R}\times S^2 $ and study the analytical properties of the charge expansion in $QED_3$ with $N_f$ two-component complex fermions \footnote{We take $N_f$ to be even in order to preserve parity and time reversal symmetry \cite{Redlich:1983kn}.}, which has been thoroughly studied in the last decades \cite{Borokhov:2002ib, Pisarski:1984dj, Appelquist:1988sr, Maris:1995ns, Nash:1989xx, Pufu:2013vpa, Braun:2014wja, Giombi:2015haa, Chester:2016wrc, Albayrak:2021xtd, Gusynin:2016som} due to its relevance for condensed matter (especially for the description of algebraic spin liquids \cite{Rantner:2002zz}) and its similarities with $QCD$ in four dimensions. In particular, we focus on the scaling dimension $\Delta_Q$ of charge-$Q$ monopole operators, which create topological disorder by acting in a given position of the space-time \cite{Borokhov:2002ib}. In general, their proliferation (occurring when the monopoles are relevant operators in the RG sense) confines the gauge field \cite{Polyakov:1975rs, Polyakov:1976fu}, but the screening produced by the fermions can elude confinement and realize conformal dynamics in the infrared above a critical value of $N_f$, $0<N_{f,\text{crit}}<10$ \cite{Appelquist:1988sr, Nash:1989xx, Gusynin:2016som, Braun:2014wja}. By virtue of the state-operator correspondence, the scaling dimensions of monopole operators equal the ground state energy of the theory on $R\times S^2$ in the presence of $4 \pi Q$ magnetic flux across $S^2$. The ground state energy can then be computed semiclassically in the $1/N_f$ expansion resulting in
\begin{equation}  \label{monoexp}
\Delta_Q = \sum_{j=-1}^{\infty} \frac{1}{N_f ^j}\Delta_j(Q) \,,
\end{equation}
where $\Delta_{-1}$ and $\Delta_0$ have been computed, respectively, in \cite{Borokhov:2002ib} and \cite{Pufu:2013vpa}. It can be shown that the large-$Q$ expansion of $\Delta_Q$ reproduces the general large-charge formula \eqref{largecharge}, including the correct value of the universal term scaling as $Q^0$ \cite{Hellerman:2015nra}. Here, the relevant eigenfunctions on $S^2$ are given in terms of \emph{monopole harmonics} \cite{Wu:1976ge, Wu:1977qk}, generalizing the spherical harmonics in the presence of the field generated by a magnetic monopole. Interestingly, the properties of these harmonics (in particular the fact that their angular momentum is bounded from below by the charge) lead to considerable differences with respect to the $O(N)$ theory on $\mathbb{R}\times S^2$. In fact, we find that the large-$Q$ expansion of $\Delta_{-1}$ is only $n!$ (and not $(2n)!$) factorially divergent, with an optimal truncation order $n_{\text{opt}} \approx Q$ and related error of order $\cO\left(e^{-Q} \right)$. Moreover, the Borel transform $\mathcal{B}[\Delta_{-1}](t)$ exhibits an infinite series of equally spaced branch points at $t = 4 \pi m \ i$, $m \in \mathbb{Z} $, and, therefore, is Borel summable. The Borel sum is given in terms of an infinite sum of modified Bessel functions of the second kind, providing an alternative expression for $\Delta_{-1}$. Notice that, due to the properties of the ground state, at the leading order in $1/N_f$ the addition of a Gross-Neveu (GN) interaction term does not affect $\Delta_Q$ \cite{Dupuis:2021flq}. Therefore, our findings trivially apply also to the critical $QED_{3}-GN$ model, which is relevant for the quantum phase transition between Dirac and chiral spin liquids \cite{He:2013wcz, He:2015vnc, Janssen:2017eeu}. 
Finally, we give our conclusions in Sec.\ref{concl}.

\section{The $O(N)$ model in $d= 3-\epsilon$} \label{3minuseps}

In this section, we consider the sextic $O(N)$ CFT in $d=3-\eps$ with the Lagrangian
 \be
{\cal L} =\frac12\partial^{\mu}\phi_i\partial_{\mu}\phi_i+\frac{g^2}{8\times3!}(\phi_i\phi_i)^3.
\ee
where $\phi_i$, $i=1,\dots, N$, transforms as a $O(N)$-vector.
This model exhibits an infrared stable fixed point at \cite{Pisarski:1982vz}
\be
\frac{g^2}{(4 \pi)^2} = \frac{2 \eps}{22+3 N} + \cO\left(\eps^2\right) \,.
\ee
Interestingly, the beta function of the $g$ coupling is non vanishing from two-loops, and the model is, therefore, conformally invariant in $d=3$ at the one-loop level. This property will allow us to directly compare our results to those in \cite{Dondi:2021buw} and link them to the large-charge effective theory describing the three-dimensional $O(N)$ CFT.
In \cite{Jack:2020wvs}, the scaling dimension of the lowest-lying operators with total charge $Q$ \footnote{It can be shown that in the perturbative regime, i.e. in absence of level-crossing, these operators transform as traceless symmetric $O(N)$ tensors and can be written as \begin{equation}
    T_Q =  t_Q^{i_1...i_Q}(\phi_i) \ ,
\label{pmldef}
\end{equation} 
where $t_Q^{i_1...i_Q}(\phi_i)$ is a fully symmetric and traceless homogeneous polynomial of degree $Q$ in the $\phi_i$'s. For instance $t^{i}_{1}(\phi) = \phi^i$ and $t^{ij}_{2}(\phi)= \phi^i \phi^j - {\frac{1}{N}} \delta^{ij} \phi_k \phi_k$. Physically, the $\Delta_Q$ control the critical behavior of $O(N)$-invariant systems subject to anisotropic perturbations, e.g. density-wave systems \cite{Brock-etal-86}, magnets with a cubic crystal structure \cite{Aharony-76}, and superconductors \cite{Zhang}.} has been computed in the double-scaling limit \eqref{doubleQeps}, where one can perform the semiclassical expansion of \eqref{TQ}.
The leading order $\Delta_{-1}$ is given by evaluating the action on the non-trivial classical trajectory induced by fixing the charge and it reads 
\be
\Delta_{-1}(g Q)=g Q F_{-1}\left(\frac{g^2Q^2}{2\pi^2}\right) \,, \quad  F_{-1}(x)=\frac{1+ \sqrt{1+x}+\tfrac x3}{\sqrt2(1+\sqrt{1+x})^{\frac32}} \,, \quad x = \frac{g^2Q^2}{2\pi^2} \ .
\label{DelmQ}
\ee
At the next-to-leading order of the semiclassical expansion \eqref{TQ}, one needs to compute the functional determinant of the fluctuations around the classical solution. This can be formally written as
\be
\Delta_0(g Q)=\Delta_0^{(a)}(g Q)+\left(\frac{N}{2}-1\right)\Delta_0^{(b)}(g Q) \,,
\label{ONDela}
\ee
with
\begin{align} 
\Delta_0^{(a)}(gQ) &=\frac{1}{2}\sum_{\ell=0}^{\infty}n_\ell[\omega_+(\ell)+\omega_-(\ell)] \,, \\ 
\Delta_0^{(b)}(g Q)&=\sum_{\ell=0}^{\infty}n_\ell \ \omega_{*}(\ell) \,.
\label{ONDel}
\end{align}
Here
\be
\omega_{\pm}^2(\ell)=J_\ell^2+2\left(2\mu^2-\frac{(d-2)^2}{4}\right)\pm2\sqrt{J_\ell^2\mu^2+\left(2\mu^2-\frac{(d-2)^2}{4}\right)^2} \,, \qquad \omega_{*}(\ell)=\sqrt{J_\ell^2+\mu^2} \,,
\label{omdef}
\ee
are the dispersion relations of the spectrum. The latter contains a massless mode $\omega_-$, (the \emph{conformal mode}), a gapped mode $\omega_+$ with mass $\omega_+(0) = 2\sqrt{2\mu^2-\frac{(d-2)^2}{4}}$ (the \emph{radial mode}) as well as $(N-2)$ gapped modes $\omega_*$ with mass $\omega_*(0) = \mu$ (the \emph{spectator modes}). The above expressions are explicit functions of the chemical potential $\mu$, which is related to the 't Hooft coupling $g Q$ through the equations of motion as \footnote{The chemical potential is measured in units of the compactification radius (which is fixed to unity) and is, therefore, dimensionless.} 
\be
\mu=\frac{1}{2\sqrt2}\sqrt{1+\sqrt{1+\frac{g^2Q^2}{2\pi^2}}} \,.
\label{mudefa}
\ee
After regularizing the sums over $\ell$ in Eq.\eqref{ONDel}, one obtains the following final expression for the functional determinants \cite{Jack:2020wvs}
\begin{align}
\label{Delab1}
\Delta_{0}^{(a)}(g Q) &=\frac14-3\mu^2+\tfrac12\sqrt{8\mu^2-1}+\frac12\sum_{\ell=1}^{\infty}\sigma^{(a)}(\ell) \,,  \\
\Delta_{0}^{(b)}(g Q) &=-\frac14-\mu^2+\mu+\frac12\sum_{\ell=1}^{\infty}\sigma^{(b)}(\ell) \,,
\label{Delab2}
\end{align}
 where
\begin{align}
\sigma^{(a)}(\ell)=&(1+2\ell)[\omega_+(\ell)+\omega_-(\ell)]
-4l(l+1)-6\mu^2+\frac12 \,, \\
\sigma^{(b)}(\ell)=&2(1+2\ell)\omega_{*}(\ell)-4l(\ell+1)-2\mu^2-\frac12 \,,
\label{sigab}
\end{align}
are convergent sums.

\subsection{The small-charge expansion}
Equipped with the basic setup above we can now study the convergence properties of the small $g Q$ expansion of $\Delta_{-1}$ \eqref{DelmQ} and $\Delta_0$ \eqref{ONDela}.  In this limit,  $\Delta_{-1}$ is convergent with a radius of convergence determined by the only non-analytical point at $x_0=-1$. \footnote{Since $x_0<0$, one can trivially analytically continue the small-charge expansion to every positive value of $x$ and continuously connect the small- and large-charge expansions for any real value of $g Q$.} It is therefore instructive to determine how many coefficients of the small $g Q$ expansion are needed in order to fully characterize the singularity. This is achieved by making use of the Darboux's theorem which links the large order behaviour of the expansion coefficients about one point (which we take to be $x=0$) to the behaviour of the function in the vicinity of its singularities. Concretely, if the perturbative coefficients of a function $\mathcal{O}(x)=\sum c_n x^n$ grow as
\begin{equation}
    c_n \sim \frac{1}{x_0^n} \left[f(x_0) \binom{n+p-1}{n}-x_0 f'(x_0)  \binom{n+p-2}{n}+ \frac{x_0^2}{2!}f''(x_0) \binom{n+p-3}{n} -\dots \right] + \dots
    \label{darbo} \, ,
\end{equation}
then $x_0$ corresponds to the closest singularity  to the origin and further determines the radius of convergence of the expansion around $x=0$. Moreover, in the vicinity of $x_0$, $\mathcal{O}(x)$ behaves as
\begin{equation}
  \mathcal{O}(x)= f(x) \left(1-\frac{x}{x_0} \right)^{-p} + \text{analytic} \,, \qquad x \to x_0 \,,
\end{equation}
with $f(x)$ an analytic function near $x_0$. Given the $c_n$, the parameters entering Eq.\eqref{darbo} can be determined by considering various sequences which tend to them in the limit $n \to \infty$ and making use of acceleration methods to improve the convergence. For instance, the ratio of consecutive coefficients $c_n/c_{n-1}$ converges to $1/x_0$ as $n \to \infty$ \footnote{The radius of convergence can be found also by considering $\lim_{n \to \infty} \rvert a_n \rvert^{-1/n} = x_0$. However, for the series considered in this paper, the simple ratio test performs better.}, whereas $p$ and $f(x_0)$ can be found by considering the following sequences
\begin{align}
    &p =1 + \lim_{n \to \infty} n \left(x_0 \frac{c_n}{c_{n-1}} - 1 \right) \,, \\
    &   \nonumber \\
    &f(x_0) =\lim_{n \to \infty} \frac{c_n}{\left(\frac{1}{x_0}\right)^n \binom{n+p-1}{n}} \,.
    \label{Eqforp}
    \end{align}
In a similar manner, one can determine all the derivatives $f^{(n)}(x_0)$ and relevant parameters characterizing the subleading singularities \cite{Darboux, Darboux2, Dorigoni:2015dha}. As summarized in Table \ref{summarynumb}, by analyzing $20$ coefficients of the small-charge expansion of $\Delta_{-1}$, we learn that they satisfy Eq.\eqref{darbo} with $x_0=-1$, $p=-3/2$,  $f(x_0)=-\frac{1}{12 \sqrt{2}}$. With the same number of coefficients, we find also that  $f'(x_0)=-0.051559(1)$, while computing higher derivatives of $f(x)$ requires an increasing number of coefficients. For instance, to obtain $f''(x_0)=-0.091150(1)$ and $f'''(x_0)= -0.2467(1)$, we had to consider $46$ and $89$ coefficients, respectively. . 
\begin{table}[t]
%\hspace{-2cm}
\begin{center}%\footnotesize
\begin{tabular}{|c|c|c|c|c|c|c|c|c|}
\hline
&\multicolumn{6}{c|}{small $Q \eps$} & \multicolumn{2}{c|} {large $Q \eps$} \\
\hline
 &\multicolumn{3}{c|}{$O(N)$ in $d=3-\epsilon$} &\multicolumn{3}{c|}{$O(N)$ in $d=4-\epsilon$} & $O(N)$ in $d=3-\epsilon$ & $O(N)$ in $d=4-\epsilon$\\ \hline
$\Delta_j$ & $\Delta_{-1}$ & $\Delta_{0}^{(a)}$ & $\Delta_{0}^{(b)}$ & $\Delta_{-1}$ & $\Delta_{0}^{(a)}$ & $\Delta_{0}^{(b)}$ & $\Delta_{-1}$ & $\Delta_{-1}$ \\
\hline
$n$ & $20$  & $>100$ & $22$ & $25$ & $>100$  & $13$ & $28$ & $36$
 \\
\hline
\end{tabular}
\end{center}
\caption{\label{summarynumb}
Number of expansion coefficients needed to determine, with a $5$ digits accuracy, the position ($x_0$), type ($p$), and amplitude $(f(x_0))$ of the leading singularity in the coefficient functions $\Delta_j$ in the $O(N)$ model in $d=3-\eps$ and $d=4-\eps$ dimensions. In the case of $\Delta_0$, we separate the contribution of $\Delta_0^{(a)}$ and $\Delta_0^{(b)}$ as in Eq.\eqref{ONDela}. In order to accelerate the convergence, we made use of the Richardson extrapolation \cite{Richpaper}, which in all cases performed better than other series acceleration methods, e.g. Shanks transforms \cite{Shanks} and Padé approximants.}
\end{table}

% \centering
% 	\includegraphics[width=0.7\textwidth]{small3minuseps.pdf}
% 	\caption{In this figure, we show the ratio of consecutive coefficients $\frac{c_n}{c_{n-1}}$ for growing $n$. The blue line represents the original coefficients, while the red, orange and green lines denote, respectively, the first three Richardson extrapolations \cite{Richpaper}. The ratio tend to the value $x_0=-1$.}
% 	\label{ratiotestsmall3}
% \end{figure}
Physically, when $x=x_0$ the chemical potential takes the value $\mu(x_0)=\frac{1}{2\sqrt{2}}$ and the mass of the radial mode $\omega_+$ vanishes in $d=3$. Therefore, as in the large-$N$ analysis of \cite{Dondi:2021buw}, the radius of convergence is dictated by the requirement of positive masses.

In order to study the small $g Q$ expansion of $\Delta_0$ we consider separately $\Delta_{0}^{(a)} = \sum_{n=0} a_n^{(a)} x^n$ and $\Delta_{0}^{(b)} =  \sum_{n=0} a_n^{(b)} x^n$. In the $\Delta_{0}^{(a)}$ case, the ratio test to determine the radius of convergence exhibits a slow convergence while the sequence \eqref{Eqforp}, fails to converge even with more than one hundred coefficients. The poor performance of the approach can be understood  simply by inspecting the $\mu(x)$ dependence of  $\Delta_{0}^{(a)}$ in Eq.~\eqref{Delab1}. Here one immediately observes the emergence of two different singular behaviour at $x_0$, one coming from the square root term (3rd term) that has a branch cut that goes like $(1+x)^{1/4}$ (corresponding to $p = -1/4$) while the rest of the expression has an expected branch cut that goes like  $(1+x)^{1/2}$  (corresponding to $p = -1/2$) . Making use of this knowledge we can accelerate the convergence process.  We conclude that the slow convergence of Eq.\eqref{Eqforp} when considering the full $a_n^{(a)}$ is due to the presence of two coincident singularities \footnote{A possible strategy to deal with the lack of convergence due to nearby or coincident singularities consists in dividing out the strongest singularity and considering the obtained coefficients \cite{Darboux}.}. Being all the terms in Eq.\eqref{Delab2}, regular in $\mu(x_0)$ the convergence of the various ratio tests is much higher in the $\Delta_{0}^{(b)}$ case as shown in Table \ref{summarynumb}. 
%and both $x_0$ and $p$ can be determined with $4$ digits accuracy with only $15$ coefficients.
We conclude that near $x_0= -1$, $\Delta_0$ behaves as
\be
   \Delta_{0}= f(x) \left(1+x \right)^{1/4}+ g(x, N) \left(1+x \right)^{1/2}+ \text{analytic} \,.
\ee

\subsection{The large-charge expansion}

As shown above, the small $g Q$ expansion of $\Delta_{-1}$ and $\Delta_0$ is convergent. Here, we move to investigate the large $g Q$ expansion, which, as we shall see, in the case of $\Delta_0$ is asymptotic and non-Borel summable. The large $g Q$ expansion of $\displaystyle{\Delta_{-1} =g Q x^{1/4}\sum_{n=-0} b_n x^{-n/2}}$ is convergent as can be seen from the ratio of consecutive coefficients, which is depicted in Fig.\ref{ratiotestsmall4}. The radius of convergence is again determined by the singularity at $x_0=-1$. 
\begin{figure}[!t]
\centering
	\includegraphics[width=0.7\textwidth]{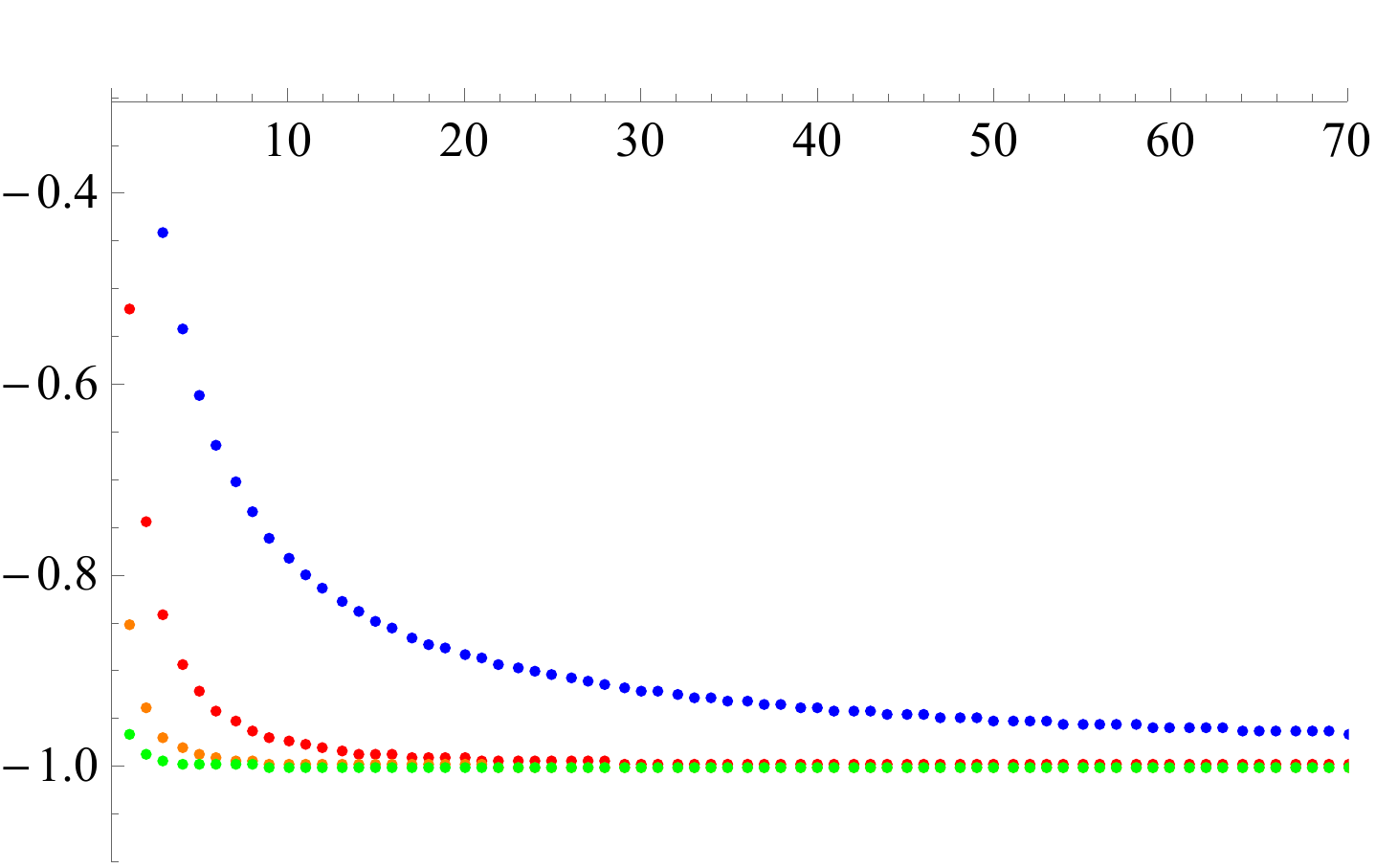}
	\caption{In this figure, we show the ratio of consecutive coefficients $\frac{b_n}{b_{n-1}}$ of the large $g Q$ expansion of $\Delta_{-1}$ for growing $n$. The blue line represents the original coefficients, while the red, orange, and green lines denote, respectively, the first three Richardson extrapolations. The ratio tends to the value $1/x_0=-1$.}
	\label{ratiotestsmall4}
\end{figure}
The number of coefficients one needs to precisely characterize the singularity is similar to the small $g Q$ case, as shown in Table \ref{summarynumb}.

%\tad{The computation of the large $g Q$ expansion of $\Delta_0$ is technically challenging and the approaches considered in the literature resorted to numerical fitting \cite{delaFuente:2018qwv, Badel:2019oxl, Badel:2019khk, Jack:2020wvs, Jack:2021ypd} and numerical evaluation of integrals \cite{Cuomo:2021cnb}, in order to determine the first few coefficients.} 

For the large $g Q$ expansion of $\Delta_0$  we focus on analytically   determining the large-charge expansion of $\Delta_{0}^{(b)}$ and leave $\Delta_{0}^{(a)}$ for future work. As we shall see later, due to the factor of $N$ in Eq.\eqref{ONDela}, our conclusions will not be affected by the inclusion of radial and conformal modes. In order to compute the large $\mu$ expansion of $\Delta_{0}^{(b)}$, we follow \cite{Jack:2020wvs} and separate the positive powers of $\mu$ as
\begin{align}
\Delta_{0}^{(b)} =a_{-3} \mu^3 + a_{-1} \mu +\sum_{l=0} (2 \ell+1) \sqrt{\mu ^2+\ell (\ell+1)}   \,.
\end{align}
The value of $a_{-3}=-2/3$ and $a_{-1}=1/3$ has been computed in \cite{Jack:2020wvs} by performing a numerical fit to $\Delta_0^{(b)}$ and will be confirmed below via an analytical computation. By Taylor-expanding the square root and exchanging the two sums, we have
\begin{align}
\Delta_{0}^{(b)} & =a_{-3} \mu^3 + a_{-1} \mu +\sum_{\ell=0} (2 \ell+1) \sqrt{\mu ^2+\ell (\ell+1)}  \nonumber \\ &= a_{-3} \mu^3 + a_{-1} \mu  +\sum_{k=0}  \frac{(-1)^k \mu^{-2 k-1} \Gamma \left(k+\frac{1}{2}\right)}{2 \sqrt{\pi } \Gamma(k+2)} \sum_{\ell=0}  (2 \ell+1) \ell^{k+1} (\ell+1)^{k+1} \nonumber \\ &=a_{-3} \mu^3 + a_{-1} \mu+ \frac{1}{\mu}\sum_{k=0} a_k \mu^{-2k} \,.
\label{giapponesi}
\end{align}
By using that
% \begin{align}
% \sum_{\ell=0} (2 \ell+1) \ell^{k+1} (\ell+1)^{k+1} &= \sum_{n=0}^{k+2} \binom{k+2}{n} \sum_{\ell=0} \ell^{k+n+1} + \sum_{n=0}^{k+1} \binom{k+1}{n} \sum_{\ell=0} \ell^{k+n+2} \nn &=\sum_{n=0}^{k+2} \binom{k+2}{n} \zeta (-k-n-1) \sum_{n=0}^{k+1} \binom{k+1}{n} \zeta (-k-n-2)  \nn &= \sum_{n=0}^{k+2} \frac{(-1)^{k+n+1} B_{k+n+2}}{k+n+2}\binom{k+2}{n} + \sum_{n=0}^{k+1} \frac{(-1)^{k+n} B_{k+n+3}}{k+n+3}\binom{k+1}{n} \,,
% \end{align}
\begin{align}
\sum_{\ell=0} (2 \ell+1) \ell^{k+1} (\ell+1)^{k+1} &= \sum_{n=0}^{k+2} \binom{k+2}{n} \sum_{\ell=0} \ell^{k+n+1} + \sum_{n=0}^{k+1} \binom{k+1}{n} \sum_{\ell=0} \ell^{k+n+2} \nn &= \sum_{n=0}^{k+2} \frac{(-1)^{k+n+1} B_{k+n+2}}{k+n+2}\binom{k+2}{n} + \sum_{n=0}^{k+1} \frac{(-1)^{k+n} B_{k+n+3}}{k+n+3}\binom{k+1}{n} \,,
\end{align}
we obtain our final expression for the coefficients 
\be
a_k = \sum_{m=1}^{k+2} \frac{(-1)^{k+1} B_{2 m} \Gamma \left(k+\frac{1}{2}\right) }{4
   \sqrt{\pi } m \Gamma (k+2)} \left [2 \binom{k+1}{-k+2 m-3}+\binom{k+1}{-k+2 m-2}\right] \,.
   \label{tricoffi}
\ee
% In App.\ref{}, we will re-obtain this result using the Euler-Maclaurin summation formula which, unlike the method used here, can be applied also to $\Delta_0^{(a)}$ and will be used in the next section to compute the large $\mu$ expansion of $\Delta_0^{(b)}$ in the $O(N)$ model in $d=4-\eps$ dimensions. 
The coefficients $a_k$ diverge double-factorially, as can be seen from the ratio $\displaystyle{\frac{a_{k+1}}{k^2 a_k}}$, which is plotted in Fig.\ref{ratiotestlarge3}. In fact,   in the $k \to \infty$ limit, they behave as
 \begin{figure}[!t]
\centering
	\includegraphics[width=0.7\textwidth]{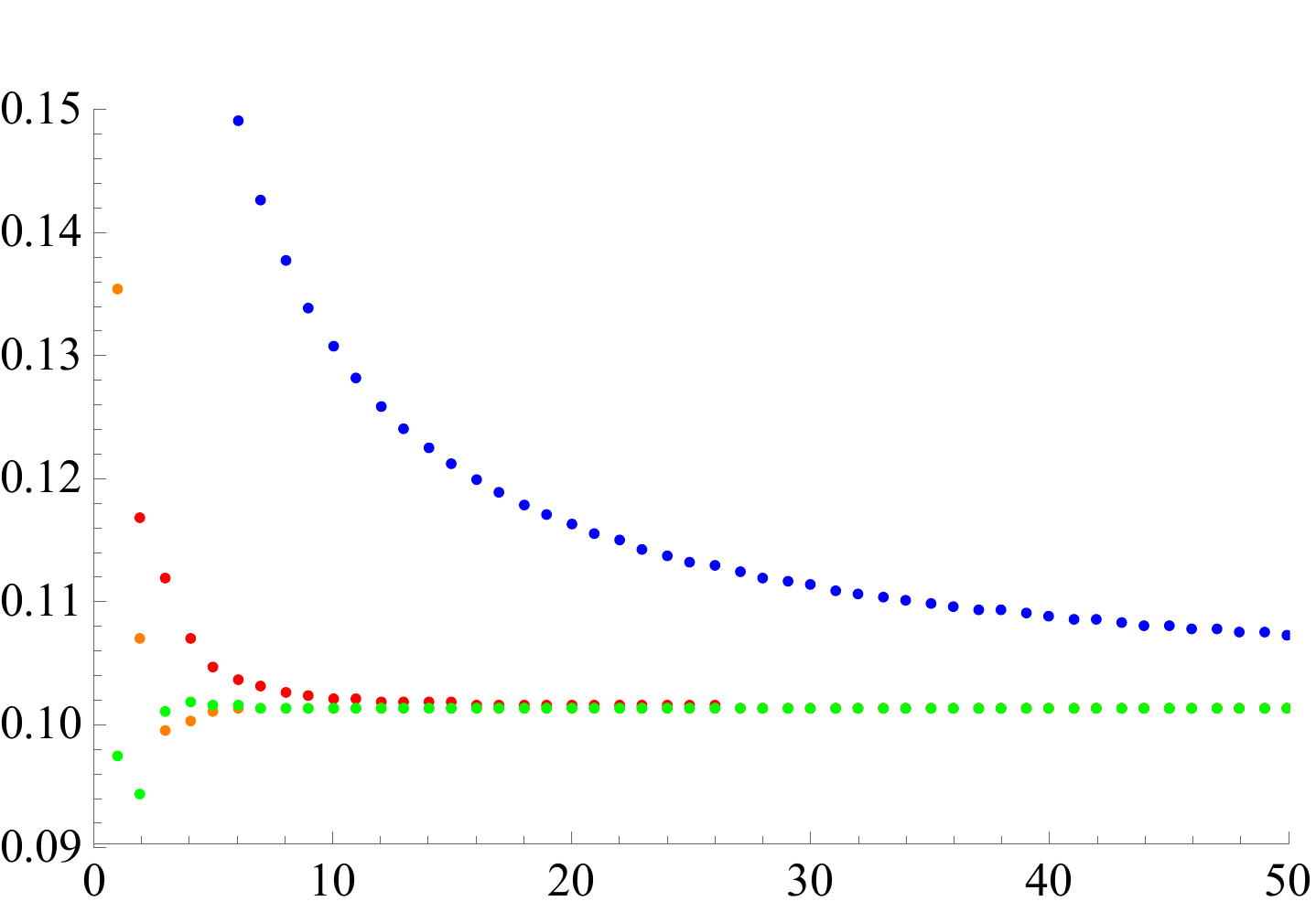}
	\caption{In this figure, we show the ratio $\frac{a_{k+1}}{k^2 a_{k}}$ (with $a_k$ given by Eq.\eqref{tricoffi}) for growing $k$. The blue line represents the original ratio, while the red, orange, and green lines denote, respectively, the first three Richardson extrapolations. The ratio tends to the value $\frac{1}{\pi^2} = 0.101321\dots$ .}
	\label{ratiotestlarge3}
\end{figure}
\be
a_k \approx -\pi ^{-2 k-5} \Gamma \left(k+\frac{1}{2}\right) \Gamma \left(k+\frac{5}{2}\right)  \,.
\ee
We can now employ resurgence arguments to infer the non-perturbative corrections to $\Delta_0^{(b)}$. According to resurgence theory, given an asymptotic series $\phi^{(0)}(z)=\sum a_k z^k$, we can promote it to a \emph{transseries} of the form
\begin{align}
    \phi(z)=\phi^{(0)}(z)+\sum_{j \neq 0} \sigma_j e^{-A_{j} / z^{1 / \beta_{j}}} z^{-b_{j} / \beta_{j}} \Phi^{(j)}(z),\qquad \Phi^{(j)}(z) \sim \sum_{i=0}^{\infty} a_{i}^{(j)} z^{i / \beta_{j}} \,,
\end{align}
where the parameters $\beta_j, A_j$ and $b_j$ are encoded in the large order behaviour of the $a_k$ coefficients as \cite{Dorigoni:2014hea, Aniceto:2018bis} 
\begin{align}
    a_{k} \sim \sum_{j} \frac{S_{j}}{2 \pi i} \frac{\beta_{j}}{A_{j}^{\beta_{j} k +b_{j}}} \sum_{i=0}^{\infty} a_{i}^{(j)} A_{j}^{i} \Gamma\left(\beta_{j} k+b_{j}-i\right) \,. \label{resurgence eq}
\end{align}
Therefore, the transseries can be mapped into the perturbative expansion up to a set of $j$-dependent constants $\sigma_j$, which are known as \emph{transseries parameters} \cite{Aniceto:2018bis}.

For our purposes, it is enough to focus on the dominant non-perturbative correction to the scaling dimension. The latter stems from the term with $m=k+2$ in  Eq.\eqref{tricoffi}. Moreover,  to ease the comparison with \cite{Dondi:2021buw}, we shift $k$ as $k \to k-2$ and introduce
\be
\hat{a}_k \equiv a_k^{(m=k+2)}\rvert_{k \to k-2}=  -\pi ^{-2 k-1} \Gamma
   \left(k+\frac{1}{2}\right) \Gamma \left(k-\frac{3}{2}\right)\zeta (2 k) \,.
\ee
To rewrite $\hat{a}_k$ in the form of Eq.\eqref{resurgence eq}, we resort to the following identity 
\begin{equation}
  \begin{aligned}
	  2^{2k} \Gamma(k+\tfrac{1}{2}) \Gamma(k - \tfrac{3}{2}) &=   \sqrt{\frac{\pi}{2} }  \sum_{i=0}^{\infty} \gamma_i \Gamma\left( 2k - \frac{3}{2} - i \right) 
  = \sqrt{\frac{\pi}{2} } \left[ 8 \Gamma(2k - \tfrac{3}{2}) + 15 \Gamma(2k - \tfrac{5}{2}) + \frac{105}{16}  \Gamma(2k - \tfrac{7}{2}) + \dots \right] \,,
\end{aligned}
\end{equation}
where the coefficients $\gamma_i$ diverge factorially and occur in Henkel's expansion of the modified Bessel function of the second kind. 
% \begin{align}
% K_2(z) \underset{z \rightarrow \infty}{\sim} \sqrt{\frac{\pi}{2z}}\ e \ a^{-z} \sum_{i =0}^\infty \left( \frac{\gamma_i}{\gamma_0} \right) \frac{1}{z^{i}} \,.
% \end{align} 
After some manipulations, we obtain
\begin{align}
\hat{a}_k=-\frac{1}{4 \pi
   ^2}\sum _{j=1} \frac{j^{-3/2}}{(2 \pi  j)^{2 k-3/2}} \sum
   _{i=0} \gamma _i \Gamma \left(2 k-\frac{3}{2}-i\right) \,,
\end{align}
which agrees with Eq.\eqref{resurgence eq} if
\begin{align}
\beta_j = 2 \,, \qquad b_j =-3/2 \,, \qquad A_j = 2 \pi j \,, \qquad \frac{S_j}{2 \pi i} a_0^{(j)} = -\frac{\gamma_0}{j^{3/2} 8 \pi^2} \,,\qquad a_{i>0}^{(j)} = \frac{a_0^{(j)}}{(2 \pi j)^i }\frac{\gamma_i}{\gamma_0} \,.
\end{align}
Taking into account the shift in $k$ performed before, we have that the dominant non-perturbative correction to $\Delta_0^{(b)}$ reads
\be \label{leadcorr}
\Delta_0^{(b)} \supset \sum_{j=1} e^{-2\pi j \mu} \mu^{3/2} \sum_{i=0} a_i^{(j)} \mu^{-i} \,.
\ee
By using Eq.\eqref{mudefa}, we can rewrite the above in terms of the charge as
\be \label{leadcorrQ}
\Delta_0^{(b)} \supset (g Q)^{5/4} \sum_{j=1}  \exp\left(-\frac{\sqrt{\pi }}{2^{3/4}}j \sqrt{g
   Q}\right)\sum_{i=0} a_i \left(2^{7/4} \sqrt{\pi} \right)^i  (g Q)^{-i/2} \,.
\ee
The leading non/perturbative contribution to the scaling dimension scales as $e^{-\sqrt{Q}}$, which is the same result obtained in \cite{Dondi:2021buw} for the three-dimensional $O(N)$ model in the double-scaling \eqref{doubleQN}. Of course, this is not surprising, since we, similarly to \cite{Dondi:2021buw}, consider the same functional determinant \eqref{labello}, whose transseries representation is unique. Below we will make this connection more precise and explicitly show how Eq.\eqref{leadcorrQ} matches the contribution of worldline instantons computed in \cite{Dondi:2021buw}, corresponding to non-trivial saddle points of the geodesic equations on the two-sphere. This can be achieved by re-deriving our results using the Mellin representation of the functional determinant of the spectator modes. We, therefore, rewrite Eq.\eqref{ONDel} as 
\be
\Delta_0^{(b)}(g Q)=\sum_{\ell=0}^{\infty}n_\ell \ \omega_{*}(\ell) = \sum_{\ell=0}^{\infty}(2 \ell + 1) \ \sqrt{\mu^2 + \ell(\ell+1)}= \frac{1}{\Gamma(s)}\int_0^\infty d t \ t^{s-1}e^{-\mu^2 t} \Tr \left(e^{\Delta_{S^2} t} \right) \Bigg \rvert_{s=-1/2} \,.
\label{KERNELPANIK}
\ee
Since in the limit $\mu  \to \infty$, the integral over $t$ is dominated by the contribution at $t = 0$, we proceed by studying the small $t$ expansion of the heat kernel $\Tr \left(e^{\Delta_{S^2} t} \right)$. By using Poisson resummation and the asymptotic expansion of the Dawson function $F(Z)$ for $z \to \infty$, we find
\begin{align}
\Tr \left(e^{\Delta_{S^2} t} \right) & = \sum_{\ell=0}^{\infty}(2 \ell + 1) \  e^{-\ell(\ell+1)t} = \frac{1}{2}\int_{-\infty}^\infty
\left| \rho \right|  d\rho  e^{-\frac{1}{4} \left(\rho ^2-1\right) t} +\frac{1}{2}\sum_{k=-\infty}^{\infty} (-1)^k \int_{-\infty}^\infty\left| \rho \right|  d\rho  e^{-\frac{1}{4} \left(\rho ^2-1\right) t+i \pi  k \rho } \nonumber \\ &
=\frac{e^{t/4}}{t}+ \sum _{k=-\infty }^\infty \frac{(-1)^k e^{t/4} \left(\sqrt{t}-2 \pi  k F\left(\frac{k \pi
   }{\sqrt{t}}\right)\right)}{t^{3/2}} = \frac{1}{t} \sum_{k=0} c_k t^k \,,
\end{align}
where
\be
c_k = \sum_{n=0}^k \frac{2^{-2 k} (-1)^{k+n+1} \left(4^{k-n}-2\right) B_{2 (k-n)}}{n! (k-n)!} \,.
\label{coff} 
\ee
By taking the integral over $t$ in Eq.\eqref{KERNELPANIK},   one recovers the correct $a_k$ coefficients \eqref{tricoffi}, including the coefficients of the positive powers of $\mu$ ($a_{-3}$ and $a_{-1}$) in Eq.\eqref{giapponesi}. Moreover, the above shows that our expansion coefficients $a_k$ in Eq.\eqref{tricoffi}, stems from the Cauchy product of the asymptotic expansion of \cite{Dondi:2021buw} with the Taylor series of $e^{t/4}$. We again focus on the leading non-perturbative correction to the heat kernel and consider
\be
c_k^{n=0} = \frac{(-1)^{k+1} \left(1-2^{1-2 k}\right) B_{2 k}}{k!} \,,
\ee
which, as expected, matches exactly the (\emph{full}) coefficients of the heat kernel expansion in \cite{Dondi:2021buw}. In particular, by using Eq.\eqref{resurgence eq}, we have that the leading non-perturbative corrections to the heat kernel have the following form
\begin{equation}
	\label{eq:non-perturbative-heat}
 \Tr \left[ e^{\Delta_{S^2} t} \right] \,\supset\, 	 2 i\left( \frac{\pi }{t} \right)^\frac{3}{2}  (-1)^{k+1} \rvert k \rvert e^{- (k\pi)^2/t} \,,
\end{equation}
which, of course, precisely matches the contribution of the worldline instantons calculated in \cite{Dondi:2021buw}. A few remarks are in order:

\begin{itemize}
    \item Due to the mismatch in $N$ of the contributions of $\Delta_0^{(a)}$ and $\Delta_0^{(b)}$ to $\Delta_0$, the non-perturbative corrections to $\Delta_0^{(b)}$ found here, survive in the full $\Delta_0$ for every value of $N$ (except obviously $N=2$ and at most another value of $N$ for which there is an exact cancellation with $\Delta_0^{a}$). In addition, there may be additional non-perturbative effects coming from $\Delta_0^{(a)}$, which may reduce the optimal truncation order below $n_{\text{opt}}=\cO\left(\sqrt{Q}\right)$.
    
 \item  Both the authors of Ref. ~\cite{Dondi:2021buw} and we start from the functional determinant of the spectator modes in $d=3$ \eqref{ONDel}. However, due to the different double-scaling limits considered, we obtain \emph{two distinct expansions}. Technically, we expanded Eq. \eqref{ONDel} in powers of $\mu$ which is the mass with respect to the Laplacian operator $\Delta_{S^2}$, which can, in turn, be expressed as a (convergent) powers series in $Q \eps$ via Eq.\eqref{mudefa}. Conversely, in \cite{Dondi:2021buw}, Eq.\eqref{ONDel} is expanded in powers of the mass $m = \sqrt{\mu^2-\frac{(d-2)^2}{4}}$ with respect to the conformal Laplacian $\Delta_{S^2}-\frac{(d-2)^2}{4}$, which, in turn, can be expressed as an (asymptotic) power series in $\frac{Q}{N}$. However, the transseries representation of $\Delta_0^{(b)}$ derived in \cite{Dondi:2021buw} via Borel resummation does not depend on such considerations and can be   obtained from Eq.\eqref{KERNELPANIK} by rewriting the heat kernel expansion as $\displaystyle{\Tr \left(e^{\Delta_{S^2}t} \right) = \frac{e^{t/4}}{t}\sum_k c_k^{n=0} t^k}$. 
    
    \item Unlike \cite{Dondi:2021buw}, where the large-charge expansion is asymptotic already at the leading order of the semiclassical expansion \eqref{doubleQN}, in our case, the $(2k)!$ factorial growth shows up only at the next-to-leading order of the semiclassical expansion \eqref{TQ}, i.e. in $\Delta_0$. In fact, due to the factor on $N$ in Eq.\eqref{ONDela}, the spectator modes contribute to the leading order of the expansion \eqref{doubleQN} and to the NLO of \eqref{doubleQeps}.
    
    \item Our results strengthen the idea that the non-perturbative effects found in \cite{Dondi:2021buw} stem from the geometry of the compactification manifold and, therefore, do not depend on the particular double-scaling limit considered. In the next section, we will, therefore, change the manifold and study the $O(N)$ model on $\mathbb{R}\times S^{3-\eps}$. This case is particularly interesting since the heat kernel on odd-spheres is known to be convergent \cite{Camporesi:1990wm}. Moreover, in Sec.\ref{sez4}, we will study the large-charge expansion in $QED_3-GN$ (Gross-Neveu) on $\mathbb{R}\times S^{2}$. Interestingly, we will show that, due to properties of the fixed-charge operators considered, the expansion is asymptotic but Borel summable.
    
\end{itemize}
\section{The $O(N)$ model around four dimensions} \label{4minuseps}

In this section, we continue analysing the convergence of the large-charge expansion in the $O(N)$ model by moving from $d=3-\eps$ to $d=4-\eps$, where we consider the renormalizable action 
\begin{equation} \label{action}
    \mathcal{S}=\int d^d x \left(\frac{(\partial \phi_i)^2}{2}+\frac{(4\pi)^2 g_0}{4!}(\phi_i\phi_i)^2\right) \,.
\end{equation}
It is well-known that this model exhibits a Wilson-Fisher infrared fixed point which is weakly coupled when $\eps \ll 1$. At the $1$-loop level, the value of the coupling at the FP reads
\begin{equation}
\label{WFFP}
g^*(\epsilon)= \frac{3 \epsilon}{8+N}+\mathcal{O}(\epsilon^2) \,.
\end{equation} 
As in the previous section, we consider the double-scaling limit \eqref{doubleQeps} and write $\Delta_Q$ as in Eq.\eqref{TQ}. The first two coefficients of the expansion \eqref{TQ} have been computed in \cite{Antipin:2020abu} (generalizing the $O(2)$ result of \cite{Badel:2019oxl}). The leading order reads
\begin{align}
 \label{classic}
  \frac{4\Delta_{-1}}{g^* Q} =  \frac{3^\frac{2}{3}\left(x+\sqrt{-3+x^2}\right)^{\frac{1}{3}}}{3^\frac{1}{3}+\left(x+\sqrt{-3+x^2}\right)^{\frac{2}{3}}}  + \frac{3^\frac{1}{3}\left(3^\frac{1}{3}+\left(x+\sqrt{-3+x^2}\right)^{\frac{2}{3}}\right)}{\left(x+\sqrt{-3+x^2}\right)^{\frac{1}{3}}} \,,  \qquad \qquad x \equiv 6 g^*  Q \,,
   \end{align}
while $\Delta_0$ is given by 
\begin{equation}
\label{eq:one-loop-det1}
\Delta_0 = \frac{R}{2}\sum_{\ell=0}^\infty n_{\ell}\left[\omega_+(\ell)+\omega_-(\ell)+(N-2)(\omega_{*}(\ell))\right]\,,
\end{equation}
where 
\begin{equation} \label{confradio}
    \omega_{\pm}(l) = \sqrt{J^2_\ell+3\mu^2-\frac{1}{4} (d-2)^2 \pm \sqrt{4 J^2_\ell\mu^2+\left(3\mu^2-\frac{1}{4} (d-2)^2\right)^2}} \,, 
\end{equation}
and
\begin{equation}
\omega_{*}(l) = \sqrt{J^2_\ell + \mu^2} \,,
\end{equation}
are the dispersion relations of the fluctuations. $J^2_\ell$ and $n_\ell$ have been given in Eq.\eqref{eigen}. The spectrum is analogous to the $d=3-\epsilon$ case, with one conformal mode $\omega_-$, one radial mode $\omega_+$ with mass $\sqrt{6 \mu^2- \frac{1}{2} (d-2)^2}$, and $(N-2)$ spectator modes $\omega_*$. Notice that the dispersion relation of the spectators does not depend on $d$ and is the same in the $d=3-\eps$ and $d=4-\eps$ cases, i.e. its functional determinant is given by Eq.\eqref{labello} evaluated in $d=4-\eps$. The chemical potential $\mu$ is related to the 't Hooft coupling $g Q$ as
\begin{equation}
  \mu = \frac{3^\frac{1}{3}+\left(x + \sqrt{-3+x^2}\right)^\frac{2}{3}}{3^\frac{2}{3}\left(x + \sqrt{-3+x^2}\right)^\frac{1}{3}} \,.
  \label{fourmu}
\end{equation}
For later convenience, we separate the contribution of the various modes as 
\be
\Delta_0(g Q)=\Delta_0^{(a)}(g Q)+\left(\frac{N}{2}-1\right)\Delta_0^{(b)}(g Q) \,,
\label{ONDela4}
\ee
where, after performing regularization and renormalization, $\Delta_0^{(a)}$ and $\Delta_0^{(b)}$ can be written in terms of convergent sums as \cite{Antipin:2020abu}
\begin{eqnarray} \label{square}
\Delta_0^{(a)}(g^* \bar Q)=  -\frac{15 \mu^4  +6 \mu^2  -5}{16}
+\frac{1}{2} \sum_{\ell=1}^\infty\sigma^{(a)}(\ell)
+\frac{\sqrt{3\mu^2-1}}{\sqrt{2}} \,,
\end{eqnarray}
\begin{eqnarray}
\Delta_0^{(b)}(g^* \bar Q)=-\frac{1}{16}\left[7 +  \mu \left(-16+6\mu+3 \mu^3\right)\right]+\frac{1}{2} \sum_{\ell=1}^\infty\sigma^{(b)}(\ell) \,,
\label{Delta0}
\end{eqnarray}
with
\begin{align}
\sigma^{(a)}(\ell) =(1+ \ell)^2 \left[ \omega_+ (\ell) + \omega_-(\ell) \right]  -2 \ell^3-6  \ell^2-2 \mu ^2-2 \left(\mu ^2+2\right)  \ell+\frac{5 \left(\mu ^2-1\right)^2}{4  \ell} \,,
\end{align}
\begin{align}
\sigma^{(b)}(\ell) &= 2 (1+ \ell)^2  \omega_* (\ell) -2 \ell^3-6 \ell^2-(\mu ^2+1)-\left(\mu ^2+5\right) \ell+\frac{\left(\mu ^2-1\right)^2}{4 \ell} \,.
\end{align}
In the following, we will unveil the large order behaviour of the small-$g Q$ and large-$g Q$ expansions of $\Delta_{-1}$ and $\Delta_0$. In particular, we will show that, when neglecting $\Delta_0^{(a)}$, both expansions are convergent as opposed to the three-dimensional case considered in the previous section.

\subsection{The small-charge expansion}

The small $g Q$ expansion of $\Delta_{-1}$ is convergent and its radius of convergence is determined by the only non-analytical point $x=x_0= -\sqrt{3}$. Notice that, being $x_0$ negative, one can smoothly connect the small- and large-charge expansions via analytic continuation. On the other hand, as observed in \cite{Antipin:2021jiw}, if one considers the model in $4-\eps$ (with $\epsilon < 0$) dimensions, where the FP occurs in the UV at negative values of $g$, then the non-analytical point lies on the positive $Q$ axis and analytic continuing to large values of $Q$  yields a complex $\Delta_Q$. The onset of complex dynamics in the large-charge sector of the quartic $O(N)$ theory above four dimensions has been previously observed in the literature. In fact, in \cite{Giombi:2020enj, Antipin:2021jiw}, it has been pointed out the existence of a critical value of the charge $Q_c$ above which $\Delta_Q$ has a non-vanishing imaginary part. In $d=4-\eps$ ($\eps<0$), and using  $x=6 g^\ast Q$  supplemented by  Eq.\eqref{WFFP} we have 
\be
Q_c \big\rvert_{1-\text{loop}}=x_0 \frac{(N+8)}{18 \eps}=-\frac{N+8}{6 \sqrt{3} \epsilon} \,,
\ee
in agreement with \cite{Giombi:2020enj, Antipin:2021jiw}.

By studying the coefficients of the small $g Q$ expansion of $\Delta_{-1}$ we have that they satisfy Eq.\eqref{darbo} with $x_0 =  - \sqrt{3}$, $p=-3/2$, $f(x_0)=\frac{1}{9}\sqrt{\frac{2}{3}}$ (obtained with $25$ terms), $f'(x_0) = 0.0014549(1)$ (with $60$), and $f''(x_0)=0.000256(1)$  (with $45$). Therefore, in the vicinity of the point $x =  - \sqrt{3}$, $\Delta_{-1}$ behaves as
\begin{equation}
   \Delta_{-1}= f(x) \left(1+\frac{x}{\sqrt{3}} \right)^{3/2}+ \text{analytic} \,.
\end{equation}
  As in the $d=3-\eps$ case, the radius of convergence occurs when the radial mode becomes massless as can be seen from Eqs.\eqref{confradio} and \eqref{fourmu}. 

%  \begin{figure}[!t]
% \centering
% 	\includegraphics[width=0.7\textwidth]{testratiosmallcharge.pdf}
% 	\caption{In this figure, we show the ratio of consecutive coefficients $\frac{c_n}{c_{n-1}}$ for growing $n$. The blue line represents the original coefficients, while the red, orange and green lines denote, respectively, the first three Richardson extrapolations. The ratio tend to the value $-\frac{1}{\sqrt{3}} = -0.5773502..$}
% 	\label{ratiotestsmall}
% \end{figure}
% We find that while only $11$ terms are needed to obtain a three digits accuracy, with $70$ coefficients the accuracy reaches seven digits. 
% The value of $p$ can be found as 
% \begin{equation}
%     p =1 + \lim_{n \to \infty} n \left(x_0 \frac{c_n}{c_{n-1}} - 1 \right)
% \end{equation}
% We find $p=-1.500(5)$ with $9$ coefficients and $p=-1.500000(5)$ with $70$ coefficients, where the parentheses denote the error on the last digit. 

% Finally $A_0$ can be determined by considering the sequence
% \begin{equation}
%     A_0 =\lim_{n \to \infty} \frac{a_n}{\left(\frac{1}{x_0}\right)^n \binom{n+p-1}{n}}
% \end{equation}
% By using $70$ coefficients we found $A_0 =0.54433106(2)$.
By investigating the small-charge expansion of the next orders in the semiclassical expansion we now test  the claim made in \cite{Badel:2019oxl}  according to which   the coefficients of the small $g Q$ expansion of $\Delta_j$, i.e. $a_{j,n}$ (i.e. $\displaystyle{\Delta_j = \sum_n a_{j,n} (g Q)^n }$ ), should obey the following large-order relation
\begin{equation}
    \frac{a_{j+1,n-1}}{a_{j,n}} \approx n \,.
    \label{ratconj}
\end{equation}
 If the above were true it would imply the following large order behaviour:
\begin{equation}
  a_{j,n} = b_j  \left(\frac{1}{-\sqrt{3}}\right)^n \binom{n+j-3/2}{n}  \left[1+ \cO\left(\frac{1}{n} \right) \right] \,,
\end{equation}
where $b_j$ are real numbers. Then, according to the Darboux's theorem, all the $\Delta_j$ would be non-analytic in $x=-\sqrt{3}$ and in the vicinity of this point would behave as
\begin{equation}
   \Delta_{j}= f_j(x) \left(1+\frac{x}{\sqrt{3}} \right)^{1/2-j}+ \text{analytic} \,.
\end{equation}
However, already for $\Delta_0$ the analysis of the coefficients of the small $g Q$ expansion  reveals that the above is incorrect. In fact, as for the case with $d=3-\eps$, near the singularity $\Delta_0$  reads
\be
   \Delta_{0}= f(x) \left(1+\frac{x}{\sqrt{3}} \right)^{1/4}+ g(x, N) \left(1+\frac{x}{\sqrt{3}} \right)^{1/2}+ \text{analytic} \,.
   \label{mahva}
\ee
In other words the arguments of \cite{Badel:2019oxl}  capture only, for $\Delta_0$, the essence of the second term in Eq.~\eqref{mahva} but not the full singularity structure. 

%In Fig.\ref{ciaociao}. we show the explicit check of Eq.\eqref{ratconj}, with $\alpha = 1$ (left) and $\alpha = 5/4$ (right). We would like to stress that, despite our findings contradicting Eq.\eqref{ratconj}, we think that the considerations made in \cite{Badel:2019oxl} are correct but can result in a more involved connection among different $\Delta_j$ than Eq.\eqref{ratconj}.  
%\begin{figure}
%\begin{center}
%\includegraphics[width=0.45\textwidth]{ratconji.pdf} $\qquad$ \includegraphics[width=0.45\textwidth]{ratconji2.pdf}
%\caption{The ratio $\frac{a_{0,n-1}}{a_{-1,n}  n^\alpha}$ as a function of $n$ for $\alpha=1$ (left) and $\alpha =5/4$ (right). The blue line represents the original coefficients, while the red, orange, and green lines denote, respectively, the first three Richardson extrapolations. Despite the right figure showing a better convergence, due to the interference of two coincident singularities in $x=-\sqrt{3}$, the analysis is inconclusive. However, progress can be made by separating the two singularities, i.e. studying separately the square root term in \eqref{square}. This procedure leads to \eqref{mahva}.}
%\end{center}
%\label{ciaociao}
%\end{figure}

Interestingly, the nature of the leading non-analytical structure characterized by $p$ in both $\Delta_{-1}$ and $\Delta_{0}$ is identical in  $d=3-\eps$ and $d=4-\eps$ dimensions for $O(N)$ theories. Intrigued by this observation, we studied the small $Q \eps$ expansion of $\Delta_{-1}$ in other two theories which have been previously investigated in the double-scaling limit \eqref{doubleQeps}, namely the cubic $O(N)$ model in $d=6-\eps$ \cite{Antipin:2021jiw} and the $U(N)\times U(M)$ model in $d=4-\eps$ \cite{Antipin:2020rdw, Antipin:2021akb}. In both cases, we find that the leading singularity $Q=Q_c$ is tied to a vanishing mass for the "radial modes" of the models.  Around this point $\Delta_{-1}$ behaves as
\begin{equation}
   \Delta_{-1}= f(Q \eps) \left(1+\left(\frac{Q}{Q_c}\right)^\beta \right)^{3/2}+ \text{analytic} \, ,
\end{equation}
where $\beta=2$ for $O(N)$ in $d=3-\eps$ and $\beta = 1$ for the other theories we investigated. The difference in $\beta$ should be traced, not in the space-time dimension, but in the fact that the model investigated in $3-\epsilon$ dimensions has one-loop vanishing beta function. %Notice that if the above is expressed in terms of critical value of $\eps$, i.e.  $\Delta_{-1} \sim f(Q \eps) \left(1+{\eps}/{\eps_c} \right)^{3/2}$, there is no parameter differentiating the models. 
Our results hint at new universal behaviours in quantum field theories.

\subsection{The large-charge expansion}

The large $g Q$ expansion of $\Delta_{-1}$ is convergent with a radius of convergence determined by the non-analytical point at $x=-\sqrt{3}$. The number of expansion coefficients needed to accurately characterize the singularity is  larger ($\approx 35$) when compared to the small $g Q$ case, as shown in Table \ref{summarynumb}.

To analyze the large $\mu$ expansion of $\Delta_0$, we focus on the contribution of the spectator fields defining $\Delta_0^{(b)}$. In particular, our goal is to prove that the large $g Q$ expansion is convergent. We use the  following Mellin representation to investigate the convergence for $\Delta_0^{(b)}$ 
\begin{align} \label{Mellano}
\Delta_0^{(b)}(g Q) &=\sum_{\ell=0}^{\infty}n_\ell \ \omega_{*}(\ell) = \sum_{\ell=0}^{\infty}( \ell + 1)^2 \ \sqrt{\mu^2 + \ell(\ell+2)}  \nonumber \\ & = \frac{1}{\Gamma(s)}\int_0^\infty d t \ t^{s-1}e^{-\mu^2 t} \Tr \left(e^{\Delta_{S^{3-\eps}} t} \right) \Bigg\rvert_{s=-1/2} = \sum_{k=0} a_k \frac{\Gamma(-1/2+k-\frac{3-\eps}{2})}{-2 \sqrt{\pi}} \mu^{4-\eps-2 k} \,,
\end{align}
where the $a_k$ are the heat kernel coefficients on $S^{3-\eps}$, i.e. $\Tr \left(e^{\Delta_{S^{3-\eps}} t} \right)\  = \sum_{k=0} a_k t^{k+\frac{3-\eps}{2}}$. For a given manifold, the heat kernel coefficients depend only on its geometrical properties, e.g. $\displaystyle{a_0 = \frac{\text{Vol.}_{S^{3-\eps}}}{(4 \pi)^{\frac{3-\eps}{2}}}}$. Due to the gamma function in the numerator of the equation above, the terms with $k=0,1,2$ diverge in the limit $\eps \to 0$ and need to be renormalized. For example, the term with $k=0$ reads
\be
-a_0 \frac{\Gamma(-2+\eps/2)}{2 \sqrt{\pi}}\mu^{4-\eps} =\left[-\frac{1}{8 \epsilon } +\frac{1}{32} (4 \gamma_E -5-4 \log (2))+ \frac{1}{8} \log (\mu )+\cO\left(\epsilon\right)\right] \mu^4 \,.
\ee
We checked that the $1/\eps$ divergence cancels against a term arising from the renormalization of $\Delta_{-1}$. As usual, the renormalization is connected with a logarithm of the relevant scale that here is given by the chemical potential. By renormalizing the first three coefficients, we obtain
\begin{align}
\Delta_0^{(b)}(g Q) &= \frac{1}{32} \mu ^4 (-5+4 \gamma_E -4 \log (2))+\frac{1}{24} \mu ^2 (1-6 \gamma_E +6 \log (2)) \nonumber \\ &+\frac{1}{80} (11+10 \gamma_E -10 \log (2))+ \frac{1}{8} \left(\mu ^2-1\right)^2 \log (\mu ) + \mu^4 \sum_{k =3} b_k \mu^{-2k}
\end{align}
in agreement with the numerical results of \cite{Jack:2021ypd}. The coefficients $b_k$ with $k\ge 3$ can be computed directly in $d=4$. The heat kernel coefficients on the $3$-sphere can be obtained as
\be
\Tr \left(e^{\Delta_{S^{3-\eps}} t} \right) = \sum_{l=0} (l+1)^2 e^{-l (l+2) t} =\frac{1}{2}e^{t}\sum_{p=-\infty}^{\infty} p^2 e^{-p^2 t} = \frac{\sqrt{\pi } e^t}{4 t^{3/2}} + \cO{\left(e^{-1/t}\right)} =t^{-3/2} \sum a_k t^k + \cO{\left(e^{-1/t}\right)} \,,
\ee
with $a_k =\frac{\sqrt{\pi}}{4 k!}$. Unlike the $d=3-\eps$ case, the heat kernel expansion has an infinite radius of convergence. By plugging the above in Eq.\eqref{Mellano}, we obtain the coefficients of the large $\mu$ expansion of $\Delta_0^{(b)}$
\be
b_{k\ge3}=-\frac{1}{4 k (k-1)(k-2)} \,.
\ee
Interestingly, we can resum the series and obtain a closed-form expression for $\Delta_0^{(b)}$ not involving infinite sums. We have
\be
\Delta_0^{(b)} = -\frac{5 \mu ^4}{32}+\frac{\mu ^2}{6} -\frac{1}{20} +\frac{1}{8} \left(\mu ^2-1\right)^2 \left(\log \left(\mu -\frac{1}{\mu }\right)+\gamma_E -\log
   (2)\right)\,.
\ee
The analytic structure of $\Delta_0^{(b)}$ is as follows: there is an essential singularity at $\mu=0$ and two logarithmic branch cuts which run, respectively, from $\mu=-1$ to $\mu=-\infty$ and from $\mu=1$ to $\mu=0$. However, from Eq.\eqref{fourmu}, we see that $\mu \neq 0$ for any value of $g Q$. Moreover, $\mu(g Q =0)=1$, and $\Delta_0$ is complex for any $Q$ when $g<0$, i.e. at the (metastable) UV FP of the quartic $O(N)$ theory in $4<d<6$. Therefore, while the small-$g Q$ expansion of $\Delta_{-1}$ reveals the existence of a critical value of the charge above which $\Delta_Q$ is complex, the analytic structure of $\Delta_0$ suggests a stronger statement, i.e. in $4<d<6$ $\Delta_Q$ is complex for \emph{any} value of $Q$. Away from four dimensions the situation can change due to different asymptotic behaviours for even and odd dimensions of the $O(N)$ CFT   \cite{Moser:2021bes}.

We have observed that the large $g Q$ expansion of the $\Delta_0^{(b)}$ is convergent, in net contrast with the $(2n)!$ factorial growth found in three dimensions. Hence our result strengthens the idea that the non-perturbative contributions to the functional determinant of spectator fields (i.e. of free particles of mass equal to $\mu$) have a geometrical origin and are, therefore, absent on $\mathbb{R}\times S^{3}$, where the WKB expansion of the heat kernel is exact \cite{Camporesi:1990wm}.

\section{Monopoles in  $QED_3$   } \label{sez4}

Here we consider the large-charge expansion in fermionic gauge theories. In particular, we study the $QED_{3}$ model with Euclidean action given by
\begin{equation}
S=\int d^{3}x\left[\frac{1}{4 e^2}F_{\mu \nu} F^{\mu \nu}+\overline{\psi}^{i}\left(\slashed{\partial} + i \slashed{A}\right)\psi^{i}
% -\frac{h^{2}}{2}\left(\overline{\psi}^{i}\psi^{i}\right)^{2}
\right] \,,
\label{eq:GN_action}
\end{equation}
where the flavor index runs over $i=1,...,N_f$ and $A_\mu$ is a $U(1)$ gauge field with field strength $F_{\mu \nu}$. The theory has a $SU(N_f)$ flavor symmetry and a $U(1)$ global symmetry associated with the current 
\be
J_\mu = \frac{1}{4 \pi} \eps_{\mu \nu \rho} F^{\nu \rho} \ ,
\ee
which is conserved due to the Bianchi identity $d F = 0$. One can define the monopole operators as the operators carrying the corresponding conserved charge $Q = \int d^2 x J_0$, which is subject to the Dirac quantization condition $Q \in \mathbb{Z}/2$.  
For large enough $N_f$, the theory is believed to flow to a conformal field theory in the infrared \cite{Appelquist:1988sr, Nash:1989xx}. In this phase we can relate the scaling dimension of the lowest-lying monopole operators to the ground state energy on the cylinder as  
\begin{equation}
    \Delta_{Q}=E_{Q} \equiv-\log Z_{S^{2} \times \mathbb{R}}\left[A^{Q}\right] \,.
\end{equation}
Here $\Delta_{Q}$ corresponds to the scaling dimension of a monopole operator carrying the charge $Q$, $E_Q$ is the ground state energy on the cylinder  when there is $4\pi Q$ units of magnetic flux across $S^2$, $A^Q$ the associated background gauge field, and  $Z_{S^{2} \times \mathbb{R}}$ is the partition function of the theory. 
For large $N_f$, $\Delta_Q$ can be computed via a semiclassical expansion in  $1/N_f$ yielding Eq.\eqref{monoexp}. The leading order corresponds to the action evaluated on the classical field configuration and reads \cite{Borokhov:2002ib, Pufu:2013vpa}
\begin{equation}
\Delta_{-1}=4 \sum_{\ell=Q+1}^{\infty} \ell \sqrt{\ell^{2}-Q^{2}} \,, \label{QED3anomalous}
\end{equation}
where $\ell$ labels the eigenvalues of the Laplacian on a $2$-sphere with a charge $Q$ at the center. The corresponding eigenfunctions are the monopole harmonics \cite{Wu:1976ge, Wu:1977qk} and the presence of the background monopole field bounds $\ell$ as $\ell \ge Q+1$. The above expression can be   regularized and computed numerically, as explained in detail in \cite{Borokhov:2002ib, Pufu:2013vpa}.

 \subsection{The large-charge expansion}
Here we focus on the large $Q$ expansion of $\Delta_{-1}$ \eqref{QED3anomalous}. By shifting the sum over $\ell$, we can rewrite $\Delta_{-1}$ as 
\begin{align}
\Delta_{-1} &= 4\sum _{n=0}^{\infty } (n+Q+1) \sqrt{(n+1) (n+2 Q+1)} = 4\sqrt{2} Q^{3/2}  \sum _{n=0}^{\infty }  \sqrt{n+1} \left(\frac{n+1}{Q}+1\right) \sqrt{\frac{n+1}{2 Q}+1} \nonumber \\ & = 4 \sqrt{2} Q^{3/2} \sum _{k=0}^{\infty } \frac{(-1)^{k-1} 8^{-k} (2 k)!
  }{(2 k-1) (k!)^2} \left[ \sum _{n=0}^{\infty } (n+1)^{k+\frac{1}{2}} \left(\frac{n+1}{Q}+1 \right) \right]  \left(\frac{1}{Q}\right)^k
 \nonumber \\ & =4 \sqrt{2} Q^{3/2} \sum _{k=0}^{\infty } \frac{(-1)^{k-1} 8^{-k} (2 k)!
  }{(2 k-1) (k!)^2} \left[\frac{\zeta \left(-k-\frac{3}{2}\right)}{Q}+\zeta \left(-k-\frac{1}{2}\right) \right]  \left(\frac{1}{Q}\right)^k \,.
\end{align}
% where we have used the Taylor expansion of the square root and the definition of the Riemann Zeta function $\xi(s)=\sum_{n=1}^{\infty}n^{-s}$. 
Rearranging the terms of the expansion, we obtain
\begin{equation}
\Delta_{-1}= Q^{3/2}\sum_{k=0}^{\infty} a_k \frac{1}{Q^k}  \,,\label{serie1}
\end{equation}
where the $a_k$ coefficients are given by
\begin{equation}
a_k = \frac{2}{ \pi^2 k!}(-1)^{k+1} \frac{1}{(4\pi)^k}  
  \Gamma \left(k-\frac{3}{2}\right)
   \Gamma \left(k+\frac{5}{2}\right)   \sin \left(\frac{\pi}{4}   (2 k+1)\right) \zeta \left(k+\frac{3}{2}\right) \,.
   \label{qedcoff}
\end{equation}
Analysing the ratio of consecutive coefficients, which we show in Fig.\ref{ratiotest}, we find that the series is asymptotic and, therefore, requires a summation prescription such as Borel resummation.
%. In particular the large order behaviour of the $a_k$ is given by\begin{equation}a_k = - 2^{-\frac{3}{2}} \pi ^{-2} \left(\frac{-1}{4 \pi} \right)^k \sin \left(\frac{1}{4} \pi  (2k+1)\right) \Gamma (k) \left(1+ \cO\left( \frac{1}{k}\right)\right) \label{largeorder}\end{equation}
% \begin{align}
% a_{k}^{(1)}&=    \frac{(-1)^{k}(2 k) !}{(1-2 k)(k !)^{2} 2 ^{3k}} \xi(-k-1/2)\\
% a_{k}^{(2)}=&    \frac{(-1)^{k}(2 k) !}{(1-2 k)(k !)^{2} 2 ^{3k}} \xi(-k-3/2)
% \end{align}
% Usually the summing procedure introduce ambiguities depending of the position of the poles in the Borel plane. 
The Borel transform of Eq.\eqref{serie1} is given by
\begin{figure}[!t]
\centering
	\includegraphics[width=0.7\textwidth]{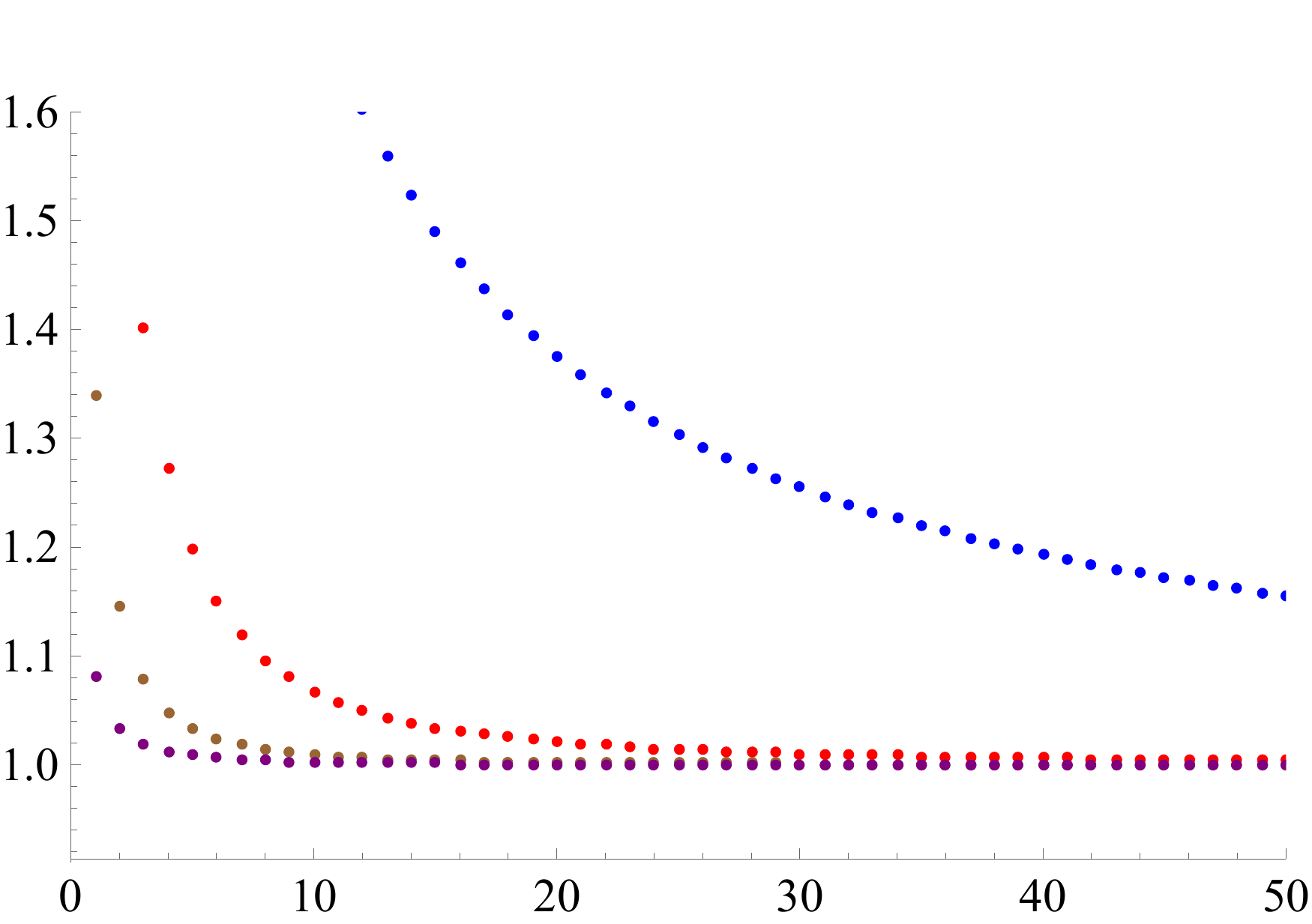}
	\caption{In this figure, we show the ratio $\displaystyle{\frac{4 \pi }{ k (-1)^{k+1}}\frac{a_k}{a_{k-1}}}$, where the $a_k$ are given by Eq.\eqref{qedcoff}. The blue line represents the original coefficients, while the red, brown, and purple lines denote, respectively, the first three Richardson extrapolations.}
	\label{ratiotest}
\end{figure}
%The Borel transform of Eq.\eqref{largeorder} reads
%\begin{equation}   \mathcal{B}[F](t)= \frac{1}{\pi}\frac{t-4 \pi}{t^2+(4\pi)^2}
%\end{equation}
\begin{equation}
\mathcal{B} \left[\frac{\Delta_{-1}}{ Q^{3/2}}\right](t)= \sum_{k=0}^\infty \frac{a_k}{k!} t^k = \sum _{m=1}^{\infty } \frac{\left(i-1\right) }{\sqrt{2} \pi  m^{3/2}} \left[\,_2F_1\left(-\frac{3}{2},\frac{5}{2};1;-\frac{i t}{4 m \pi }\right)+i \ 
   _2F_1\left(-\frac{3}{2},\frac{5}{2};1;\frac{i t}{4 m \pi }\right)\right] \,.
   \label{boring}
\end{equation}
Here $\,_2F_1(a, b; c; x)$ denotes the Hypergeometric function, which can be analytically continued in the complex plane along any path avoiding the branch points at $x=1$ and $x=\infty$. Hence $\mathcal{B} \left[\frac{\Delta_{-1}}{Q^{3/2}}\right](t) $ features an infinite series of branch points at $t=4\pi i m$, $m \in \mathbb{Z}$, as shown in Fig.\ref{polesQED3}.
\begin{figure}[!t]
\centering
	\includegraphics[width=0.7\textwidth]{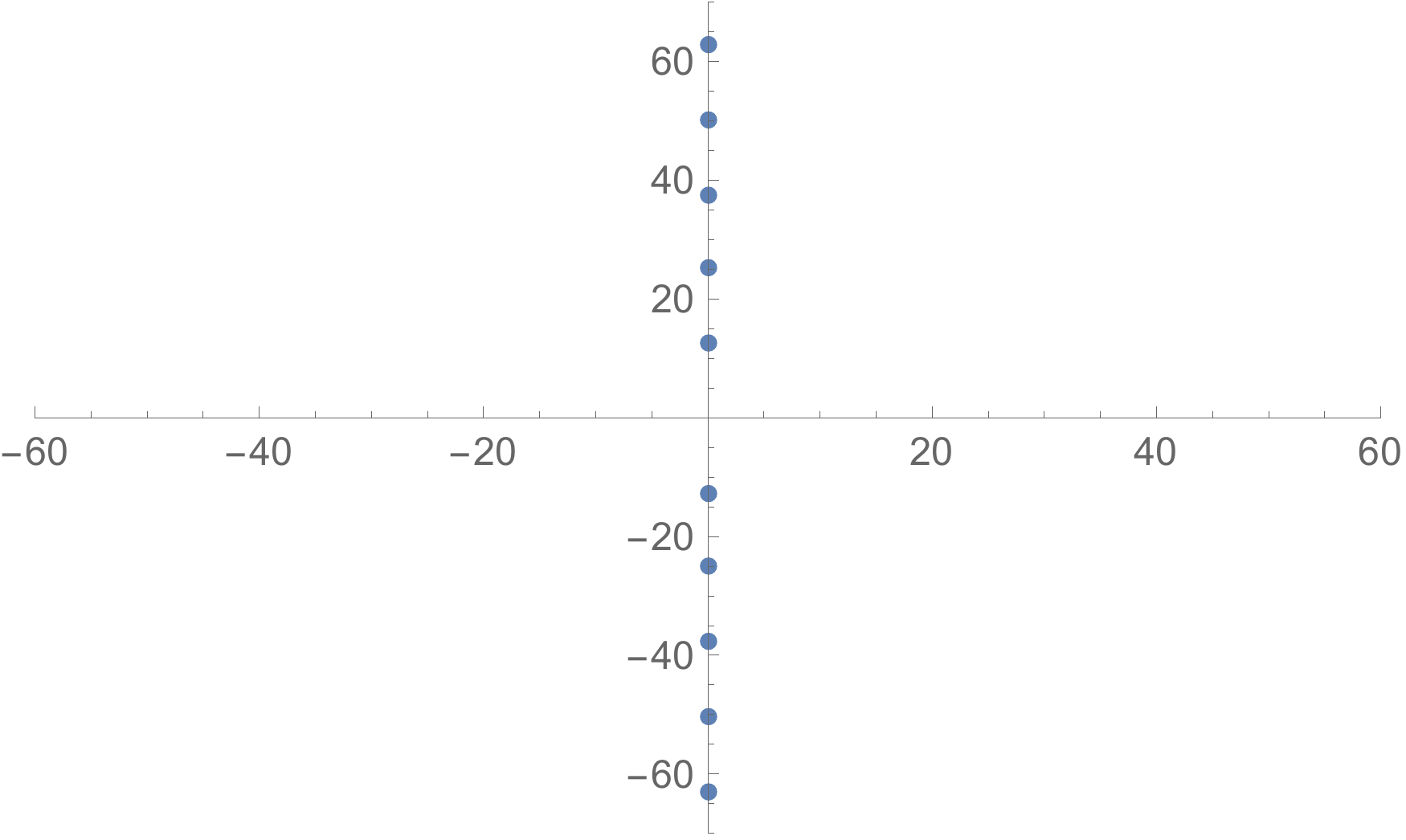}
	\caption{The singularity structure of the Borel transform of $\Delta_{-1}$ \eqref{boring}. There are branch points at $t=4\pi i m$, $m \in \mathbb{Z}$.}
	\label{polesQED3}
\end{figure}
As a consequence, the series \eqref{serie1} is Borel summable and both lateral Borel summations coincide
\begin{align}
\Delta_{-1}= Q^{5/2} \int_0^{\infty} d t \ e^{- Q t} \mathcal{B} \left[\frac{\Delta_{-1}}{ Q^{3/2}}\right](t)  =\sum_{m=1} \frac{2 i Q^2 }{\pi  m} \left[e^{x} K_2(x)-e^{-x} K_2(-x)\right]  \,, \qquad x \equiv 2 i \pi  m Q \,,
\end{align}
where $K_{2}$ is the modified Bessel function of the second kind. Finally, the optimal truncation order corresponds to the value of $k$ such that $a_k/Q^k$ has a minimum and reads
\be
k_{\text{opt}} \approx 4 \pi Q \,,
\ee
with an error of order $\cO\left(e^{-4 \pi Q}\right)$. We conclude that, even if the $QED_3$ model shares the same universal large-charge behaviour \eqref{largecharge} of the three-dimensional $O(N)$ model, its large-charge expansion behaves better than $O(N)$, having a higher optimal truncation order (i.e. $\approx Q$ rather than $\approx \sqrt{Q}$) and being Borel summable. 
Notice that, since at the leading order in $1/N_f$ the scaling dimensions are not affected by the inclusion of a Gross-Neveu interaction term, our results apply also to $QED_3-GN$ \cite{Dupuis:2021flq}. 

\section{CONCLUSIONS} \label{concl}

In this work we studied the analytic structure of the fixed charge expansion for $O(N)$ in different space-time dimensions and $QED_3$. We have seen that in $d=3-\epsilon$ dimensions the contribution to the $O(N)$ fixed charge conformal dimensions, obtained in the double scaling limit of large charge and vanishing $\epsilon$, is non-Borel summable. Additionally, we have shown that the series is doubly factorial divergent and displays $\sqrt{Q}$ optimal truncation order. Resurgence technologies helped us show that the singularities in the Borel plane are connected to worldline instantons that were found in the alternative double scaling limit of large $Q$ and $N$ of Ref.~\cite{Dondi:2021buw}. We have also explored the case of $d=4-\epsilon$ and shown that in the same large $Q$ and small $\epsilon$ regime the next order corrections to the scaling dimensions amount to a convergent series. The resummed series exhibits a new branch cut singularity which we found to be relevant for the stability of the large charge sector of the $O(N)$ model for negative $\epsilon$. In the future, it would be interesting to include the contribution of radial and conformal modes to learn how they affect the analytic structure of the fixed charge expansion. For the $QED_3$ model we discovered that at leading order in the large number of matter field expansion the large charge scaling dimensions are Borel summable, single factorial divergent and with order $Q$ optimal truncation order. It would be also interesting to investigate whether a non-Borel summable  expansion emerges at subleading $1/N_f$ orders ($\Delta_0$ has been computed in \cite{Pufu:2013vpa} for $QED_3$ and in \cite{Dupuis:2021flq} for $QED_3-GN$).

\section*{ACKNOWLEDGEMENTS}

The work of O.A. is partially supported by the Croatian Science Foundation (HRZZ) project “Heavy hadron decays and lifetimes” IP-2019-04- 7094. M. T. was supported by Agencia Nacional de Investigación y Desarrollo (ANID) grant 72210390.

  \end{document}